\documentclass[sigconf]{acmart}

\usepackage{booktabs} 
\usepackage[utf8]{inputenc}
\usepackage{enumitem}
\usepackage{dirtytalk}
\usepackage{amsmath}

\copyrightyear{2019} 
\acmYear{2019} 
\setcopyright{acmcopyright}
\acmConference[GIR'19]{13th Workshop on Geographic Information Retrieval}{November 28--29, 2019}{Lyon, France}
\acmBooktitle{13th Workshop on Geographic Information Retrieval (GIR'19), November 28--29, 2019, Lyon, France}
\acmPrice{15.00}
\acmDOI{10.1145/3371140.3371142}
\acmISBN{978-1-4503-7260-2/19/11}

\setcopyright{rightsretained}

\acmConference[GIR'19]{The 13th Workshop on Geographic Information Retrieval}{November 2019}{Lyon, France} 
\acmYear{2019}
\copyrightyear{2019}

\begin{document}
\title{Spatial Search Strategies for Open Government Data: A Systematic Comparison}



\author{Auriol Degbelo} 
\orcid{0000-0001-5087-8776}
\affiliation{%
  \institution{Institute of Geography, Osnabrück University}
  \streetaddress{Seminarstrasse 19 a/b}
  \city{Osnabrück} 
  \country{Germany} 
  \postcode{49074}}
\affiliation{\institution{Institute for Geoinformatics, University of Münster}
  \streetaddress{Heisenbergstrasse 2}
  \city{Münster} 
  \country{Germany} 
  \postcode{48161}}  
\email{auriol.degbelo@uni-osnabrueck.de}

\author{Brhane Bahrishum Teka} 
\affiliation{%
   \institution{Institute for Geoinformatics, University of Münster}
  \streetaddress{Heisenbergstrasse 2}
  \city{Münster} 
  \country{Germany} 
  \postcode{48161}
}
\email{brhanebt01@gmail.com}

\renewcommand{\shortauthors}{Degbelo and Teka}

\begin{abstract}
The increasing availability of open government datasets on the Web calls for ways to enable their efficient access and searching. There is however an overall lack of understanding regarding spatial search strategies which would perform best in this context. To address this gap, this work has assessed the impact of different spatial search strategies on performance and user relevance judgment. We harvested machine-readable spatial datasets and their metadata from three English-based open government data portals, performed metadata enhancement, developed a prototype and performed both a theoretical and user-based evaluation. The results highlight that (i) switching between area of overlap and Hausdorff distance for spatial similarity computation does not have any substantial impact on performance; and (ii) the use of Hausdorff distance induces slightly better user relevance ratings than the use of area of overlap. The data collected and the insights gleaned may serve as a baseline against which future work can compare.
\end{abstract}

%
%



\maketitle

\section{Introduction}
\label{sec:introduction}
Thanks to a growing number of countries committing to open data principles, an increasingly large amount of open government datasets is currently available on the Web. Many of these datasets are georeferenced (and in the absence of reliable statistics, extrapolating results from previous work \cite{Hahmann2013} suggests that at least 60\% could be georeferenced). Within GIScience, re-use of these datasets has attracted the interest from research, and previous work has suggested, inter alia, semantic application programming interfaces to retrieve datasets according to their thematic categories \cite{degbelo2016designing}, a platform to monitor open data re-use \cite{degbelo2020tell}, a one-stop portal for open data search \cite{Hinz2018a}, and a vision of intelligent geovisualization to exploit these datasets \cite{degbelo2018intelligent}. Direct and indirect costs of open geospatial data provision were discussed in \cite{Johnson2017}; \citet{benitez2018roadblock} presented an empirically-derived taxonomy of barriers to open data re-use from a user's standpoint; and \citet{Benitez-Paez2018} proposed a conceptual framework to improve the reusability of open geographic data in cities.

Despite these early achievements, much work is still needed to help users take advantage of existing open government datasets (OGD). One area deserving more attention, in particular, is that of search of these datasets. The awareness that datasets are peculiar enough to deserve their own treatment in information search has led Google to introduce Google Dataset Search \cite{Noy2019}. More specifically, in the context of open government, \citet{Zuiderwijk2016a} report a `lack of search support' for OGD. This issue is echoed by \citet{Koesten2017}, who pointed out that finding data is a major issue for data practitioners, and the information they need to evaluate their fitness of use is not always available or easy to interpret out of context. Along the same lines, \citet{Xiao2019} identified finding usable (i.e. what they call `content-relevant') OGD as a current major challenge. 

An example of work underway to improve the discoverability of georeferenced OGD is \citet{lafia2018improving}. Their work tackled the problem of heterogeneous naming of semantically similar content through the addition of semantic annotations to metadata. This article intends to provide another take on the OGD discoverability issue. In particular, the main goal is to shed some light on the merits of different spatial search strategies for OGD. The research question investigated is: \textit{What is the impact of different spatial search strategies on performance and user relevance judgment?} Putting the question under scrutiny has both theoretical (i.e. an understanding of aspects of search that boost desirable outcomes) and practical (i.e. recommendations for designers of search systems) value. The key contribution of the article is insights from the systematic comparison of 11 strategies for spatial search. \textcolor{black}{A byproduct of the investigation is a series of methodical steps to holistically assess the merits of spatial search strategies more broadly}.

The search strategies were chosen to examine four aspects of OGD retrieval more closely: (i) the impact of space as extra dimension; (ii) the impact of the spatial similarity function (i.e. area of overlap or Hausdorff distance);  (iii) the impact of the query expansion approach (i.e. synonyms only vs synonyms, hypernyms, and hyponyms); and (iv) the impact of the query expansion source (i.e. WordNet or ConceptNet). Related work is presented briefly in Section \ref{sec:related}, touching on the topics of open government data, relevance in geographic information retrieval (GIR) and query expansion. The method used during the study and the 11 strategies examined are introduced in Section \ref{sec:method}. Section \ref{sec:theoretical} presents the results of a theoretical evaluation assessing the performance of the search strategies, and Section \ref{sec:userstudy} reports on a user study with 16 participants assessing the impact of the strategies on user relevance judgment. The implications of all results are discussed in Section \ref{sec:discussion}, and Section \ref{sec:conclusion} concludes the article.

\section{Related Work}
\label{sec:related}
As said above, this work systematically compares the impact of spatial search strategies on performance and relevance judgment in OGD. To set the scene, this section introduces previous work on OGD more broadly, relevance in geographic information retrieval, and query expansion (four search strategies use query expansion as a technique). Overall, the key takeaway is that research on OGD is vibrant, yet our understanding of `user relevance' in the context of georeferenced OGD items is still limited.

\subsection{Open Government Data}
With the emergence of open government data portals around the world in the last decade, open government data has become a burning research area. As a result various works have attempted to address open government data from different perspectives including but not limited to OGD management, policies, legal issues, usage and values (in social sciences), OGD infrastructure and interoperability, cleaning, quality assessment, visualization, linking, publishing, mining, rating and feedback methods (in information sciences) \citep{Charalabidis2016}. As the main focus of this research is on data discovery, this section briefly presents previous work on linked open government data, OGD spatial information retrieval, as well as OGD platforms and working principles.

\subsubsection{Linked Open Government Data (LOGD)}
Despite governments’ reduced cost of providing data to consumers thanks to open government data portals, making OGD datasets available as raw datasets has made the human workload of making them machine-understandable bigger. The need for effective infrastructure, therefore, arises from the necessity of distributing this workload and facilitate easier use of government data by the community of users and developers \citep{Ding2011}. The Semantic Web and linked open government data (LOGD) \citep{Ding2010} overcome provision, reuse, and integration limitations by exposing OGD as interlinked datasets to the public via RDF (Resource Description Framework) and SPARQL endpoints. This allows users and developers to access linked data in JSON and XML and easily build applications that make use of LOGD.

\citet{Ding2011} developed a Semantic Web-based \href{https://logd.tw.rpi.edu}{LOGD Portal} to facilitate the usage of LOGD, increase the reuse of data and thereby serve the growing international community of open government data. This work has been used as a base for data.gov by converting the datasets in data.gov into RDF and then again back to data.gov to enable users easier discovery of open data and relationships between these data. \citet{Rozell2012}, on the other hand, developed an International Open Government Data Search system that performs information retrieval on a catalog of open government datasets aggregated from 43 countries. The system allows users to filter datasets by keyword terms from titles and descriptions, source catalogs, countries of origin, category tags and so on.

\subsubsection{OGD Spatial Information Retrieval}
Although the LOD works are relevant regarding open government data portals, data reuse and give insight into the goal and research focus of OGD, they do not tackle the spatial information retrieval issue approached in this work. More related to our work, \citet{Neumaier2019} proposed algorithms to add spatio-temporal annotations to datasets from OGD portals, as a first step for OGD retrieval based on spatio-temporal properties. \citet{kuo2019metadata} proposed a scoring of OGD metadata based on their spatial and temporal properties. \citet{DeFernandesVasconcelos2017} proposed to improve spatial queries in open government data portal by performing spatial similarity of area of overlap based on bounding box and ranking at a resource level. They evaluated their approach with the help of harvested datasets from OGD Brazil using a CKAN API. Their work only considers spatial search without due consideration to thematic queries that we incorporated in this study. Another set of works \citep{Chen2018,Lacasta2017,Jiang2018} was mainly about spatial data infrastructures or geospatial catalogs based on metadata, but is of interest for our study.  

\citet{Lacasta2017} proposed discovering related geospatial data in different resources by taking all metadata records of resources that partially fulfill a query (i.e., intersect the bounding box or only match the themes), find their spatial \& thematic relations and generate sets of metadata records or results that are a better answer to the query than each one individually. Even though their work is not exclusively for open government data, it was of interest for our research work for three reasons. In their implementation they used Hausdorff distance as a way of result ranking which they deemed as appropriate for ordering geometries of different size like country vs region. They also reported their inability to perform simultaneous spatial and thematic search in data.gov.uk. Third, they pointed out the absence of ontologies for query expansion in such data portals. These directions were also considered in our systematic comparisons. In another work dealing with geographic information retrieval and ranking in spatial data infrastructures, \citet{Chen2018} proposed using artificial neural networks to learn from knowledge of experts to integrate the characteristics of geospatial data to the computation of an overall similarity score. Among the similarities integrated, one is thematic similarity in which they used WordNet similarity methods. They stated to have achieved a higher precision in terms of similarity computation of geospatial data but pointed out the availability of limited Geoscience related vocabularies in WordNet and the need for continuous similarity results. The use of an advanced knowledge base that improves WordNet and Hausdorff distance in this study was, therefore, motivated from their work. Finally, \citet{Jiang2018}'s conducted two studies on the topic of spatial data infrastructure. They first developed a system that improved search experience of users from oceanographic data by utilizing data relevancy from user behavior using semantic query expansion and machine learning-based ranking. They also attempted to incorporate users’ multidimensional preferences by identifying spatial similarity and metadata attributes and thereby improve the optimal user experience.

\subsubsection{Open Government Data Portal Platforms}
Different platforms are being used as open data solutions including CKAN, DKAN, Socrata, Junar\footnote{DKAN is an open data cataloging, publishing, and visualization platform by \href{https://civicactions.com/}{CivicActions} allowing governments to easily publish data to the public. Socrata is an open data platform hosting corpus of government datasets accessible via opendatanetwork.com and an API opening it up for automated exploration and research. Junar is also an early leader in Open Data publishing which offers a Software-as-a-Service (SaaS) hosting model with a fully-fledged infrastructure of hardware, software, and storage.} \citep{7460350}. Most popular government data portals nowadays are based on CKAN, the world’s leading open-source data portal \citep{OpenKnowledgeFoundation2009}. CKAN (Comprehensive Knowledge Archive Network) is a web-based management system developed by the Open Knowledge Foundation and is being used by more than 192 governments, institutions, and other worldwide organizations to manage open government data \citep{OpenKnowledgeFoundation2009}. The popular open government data portals reviewed in this work like the \href{europeandataportal.eu}{European Data Portal}, \href{data.gov}{Data.gov}, \href{data.gov.uk}{Data.gov.uk}, \href{data.gov.ie}{Data.gov.ie} are also based on CKAN. CKAN provides RESTFUL APIs for data access that were used for harvesting in this work \citep{OpenKnowledgeFoundation2009}. In open government data portals in general (and CKAN in particular) full-text search is one integral functionality. Full-text search, defined in \citep{Klc2016}, as the ability to query and possibly rank documents based on relevance, in its simplest form helps us to find documents containing given keywords ranked by their frequency in the document. CKAN is written in Python and uses Solr, a Java-based open-source information retrieval library, to achieve full-text search functionality on the datasets stored in it's PostgreSQL backend \citep{Targett2015a}. However, Solr is not the only popular information retrieval alternative \citep{Klc2016}. Despite the use of external libraries on top of PostgreSQL for full-text search, PostgreSQL also provides full-text functionality on its own, which is powerful enough for simpler applications \citep{PostgreSQLGlobalDevelopmentGroup2016,Belaid2015}. PostgreSQL's full-text search uses pre-processing and indexing to prepare documents and save for later rapid searching. The pre-processing is done by breaking documents into words, removing stop words, converting words into lexemes, optimizing and storing the preprocessed documents \citep{PostgreSQLGlobalDevelopmentGroup2016}. While the storage of pre-processed queries is done using vector datatype (tsvector), tsquery is used for making preprocessed queries \citep{PostgreSQL2018}. Both tsvector and tsquery are used in this work for full-text searching and ranking.

\subsection{Query Expansion}
Another relevant topic for this study is query expansion. Query expansion has been proposed as a way of improving search results by adding expansion terms to users' search keywords \citep{Azad2017}. The source of query expansion terms is critical in query expansion studies. Knowledge bases like WordNet are of high interest in this case because they are built manually by experts and are regarded as highly accurate \citep{Pal2013}. On the other hand, WordNet has a lower coverage of geospatial keywords \citep{Chen2018}. Since our work considers open government data with potential geospatial terms, another knowledge base (ConceptNet) was considered as an alternative. Despite the difficulty of using concepts as query expansion terms, ConceptNet has been evaluated for query expansion by \citep{Hsu2008, Rivas2014, Hsu2006, Bouchoucha2013, Azad2017}. We take a deeper look at both knowledge bases in the next subsections. 

\subsubsection{WordNet}
WordNet is a linguistic database of English words (nouns, verbs, adverbs, adjectives) organized into synonyms, which in turn denote an underlying linguistic concept \citep{Miller1990}. The following relations are represented in WordNet:
\begin{itemize}
    \item The basic relation in WordNet is synonymy. Synonyms are words that refer to the same concept and are interchangeable in many contexts. E.g: communities and residential areas or residential districts (residential area and residential district are both synonyms of community);
    \item Hyponymy and Hypernymy represent sub-type and super-type relations between synsets (i.e. unordered sets which groups synonyms together). E.g: learning and education (direct Hyponym and Hypernym), village and community (Hyponym), community and people (Hypernym);
    \item Meronymy and holonymy (part-whole), antonymy (opposites) and troponymy (which indicates manners) are additional relations used to represent semantic relations in WordNet.
\end{itemize}

\subsubsection{ConceptNet}
\href{http://conceptnet.io}{ConceptNet}, on the other hand, is a multilingual knowledge base consisting of over 1.6 million facts spanning the spatial, physical, social, temporal, psychological and other aspects of life. ConceptNet was generated from 700000 sentences from the Open Mind Common Sense Project — a collaboration of over 14000 authors. It is designed to help computers understand words expressed in natural language and consists of knowledge from sources such as Wiktionary, OpenCyc and Multilingual WordNet \citep{Speer2012RepresentingGR, MIT2019}. ConceptNet's structure is mainly made up of edges and relations. An edge (or assertion) denotes a unit of knowledge representation in ConceptNet and a relation captures relationships among edges \citep{Speer2012RepresentingGR}. ConceptNet has several types of relations, for instance \citep{Relation49:online}:

\begin{itemize}
    \item Synonym - Represents edges of similar meanings. This is the synonym relation in WordNet as well. E.g: sunlight and sunshine;
    \item IsA - A subtype or specific instance. It corresponds to the hyponym relation in WordNet. E.g: car IsA vehicle; Chicago IsA city;
     \item MannerOf - similar to IsA but for verbs. E.g: auction and sell.
\end{itemize}
    
Additional relations modelled within ConceptNet include RelatedTo, FormOf, PartOf, HasA, UsedFor, CapableOf, Causes, Antonym, UsedFor, DerivedFrom, SymbolOf, DefinedAs, Entails, SimilarTo. In our study, after testing a combination of WordNet relations for different queries only synonyms, hyponyms and hypernyms were leading to changes in the results so only these three relations were considered in the experiments. ConceptNet's synonym, IsA and MannerOf relations are the relation types that correspond to WordNet's synonyms and hyponyms/hypernyms respectively with 100\% certainty \citep{Relation49:online}.

\subsection{Relevance in GIR}
\label{subsec:relevance}
As pointed out in \cite{Frontiera2008}, relevance is a concept that encapsulates a relationship between a user's information need and a resource that is available to meet that need. In practice, relevance is computed as a similarity function between a representation of the user's query and representations of available information resources to answer the query. Within GIR, a distinction is drawn between `geographic relevance' and `relevance of geographic documents'. Geographic relevance refers to the ``the relevance of a geographic entity, given a specific context of usage'' \cite{DeSabbata2012}, whereas relevance of geographic documents estimates how well a given document is likely to fulfill a user's spatial information need (see \cite{Purves2018}). Relevance is used in this work in the latter sense. 

Relevance ranking has been examined for georeferenced videos in \cite{Ay2010,fritze2017feature}. \citet{Ay2010} proposed algorithms for ranking of videos based on their spatial and temporal properties, and \citet{fritze2017feature} reported on 12 criteria suggested by users to determine the relevance of georeferenced videos. Despite these works, and work examining geographic relevance (e.g. \cite{DeSabbata2012,Reichenbacher2016}), ``relevance in GIR is [still] not well defined, above all when the user is considered and not only a query in its broader context'' \cite{Purves2018}. Acknowledging the difficulty of clearly pinpointing what relevance is, \citet{Mizzaro1998} suggested to model relevance as a point in a four-dimensional space, the values of each dimension being:
\begin{subequations}
\begin{align}
& InfRes = \{Document, Surrogate, Information\};\\
& Repr = \{RIN, PIN, Request, Query\};\\
& Time = \{t(rin_0), t(pin_0), t(r_0), t(q_0), t(q_1),..., t(q_n), t(f) \};\\
& Comp = f \{Topic, Task, Context, ..., (Topic, Task, Context)\}
\end{align}
\end{subequations}

The first dimension (\textit{InfRes}) represents the set of information resources the user has access to. This may be a document (i.e. physical entity that the user will obtain after seeking the information), a surrogate (i.e. a representation of a document, consisting of elements such as title, list of keywords, abstract, author names), or some information (i.e. the non-physical entity that the user  creates when reading a document). The second dimension (\textit{Repr}) is the representation of the user's problem. Here the user has a real information need (RIN) to which they associate a perceived information need (PIN). The PIN is an implicit representation of the problematic situation in the user's mind. The user then expresses the PIN in a request (i.e. a representation of the PIN in human language) and eventually formalizes the request in a query (i.e. a representation of the request in a system language). The third dimension (\textit{Time}) refers to the set of the time points from the arising of the user's RIN until its satisfaction. Finally, the fourth dimension (\textit{Comp}) refers to three further components that can be used in the classification of different kinds of relevance, namely: topic (i.e. subject area interesting for the user); task (i.e. activity that the user will execute with the retrieved documents); and context (i.e. everything affecting the way the search takes place and the evaluation of results, but not pertaining to topic and task). The type of relevance examined in this article is of the form: \textit{rel = (surrogate, request, (topic, task, context))}. A detailed explanation of each of the components (i.e. surrogate, request, and so on) is provided in Section \ref{sec:userstudy}. 

\section{Research Method}
\label{sec:method}
As stated in Section \ref{sec:introduction}, this work intends to examine the impact of different spatial search strategies on performance and user relevance judgment. User relevance judgement was formally specified in Section \ref{subsec:relevance} above. Contrary to \cite{Frontiera2008} where relevance judgments were created before the study, relevance judgments in this case are collected \textit{during} the study in the context of a user-based evaluation. Performance in the work was assessed through two proxy measures: the number of results returned (a possible indicator of diversity), and the completion time. 

\begin{figure*}
  \frame{\includegraphics[width=\textwidth]{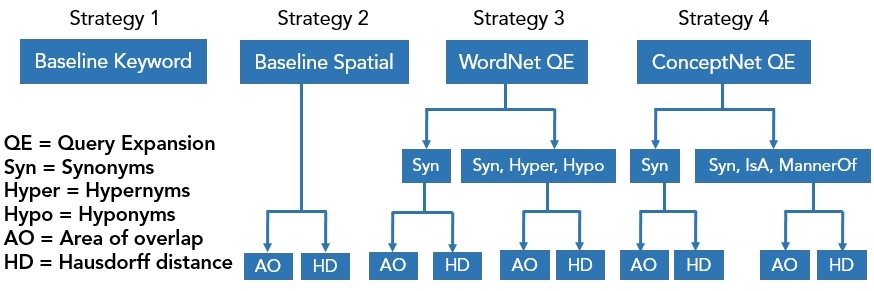}}
  \caption{Search strategies examined during the work.}
  \label{fig:Strategies}
\end{figure*}

As Figure \ref{fig:Strategies} shows, four top-level strategies were considered at the beginning of the study. The first strategy, baseline strategy, is based on only full-text search while the other strategies consider spatial search. The baseline strategy is similar to the current working approach in Data.gov.uk and Data.gov.ie (except that in this study we performed queries like (theme AND space) instead of (theme OR space)). In the second strategy, baseline spatial, location names are parsed and geocoded for spatial search. Therefore, a simultaneous spatial and thematic search is realized in this strategy. The second strategy addresses the missing simultaneous thematic and spatial search functionality in the aforementioned OGDs. The third strategy improves the second strategy by expanding the thematic keywords with expansion terms from WordNet and then applies weighted simultaneous full-text search and spatial search, followed by spatial ranking. This strategy is used to assess the impact of query expansion in OGD. Finally, the fourth strategy improves again the second strategy but by using ConceptNet for query expansion instead of WordNet and applies simultaneous weighted full-text search and spatial search and ranking. The query expansion is done using two different paths. The first uses only synonyms, while in the second we used a combination of synonyms, hypernyms, and hyponyms. The same is true of the ConceptNet-based query expansion. First we only considered ConceptNet synonyms and then a combination of synonyms, IsA and MannerOf which correspond to synonyms, hypernym, hyponym relations in WordNet. Therefore, the strategies compared totaled 11. 

\begin{figure*}
 \frame{\includegraphics[width=\textwidth]{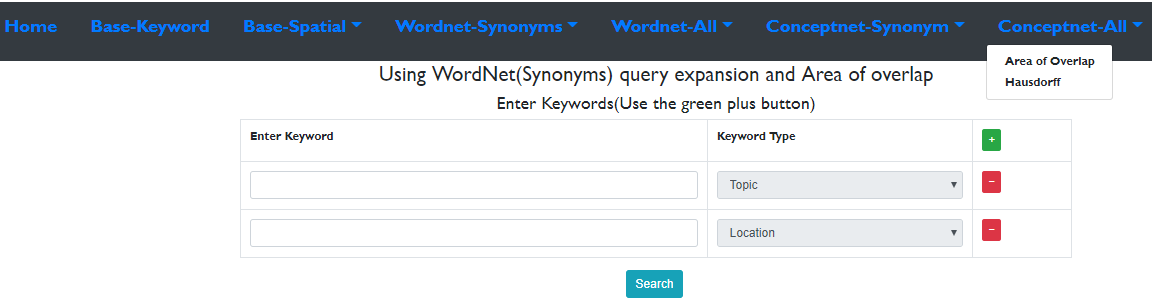}}
  \caption{User interface of the prototype.}
  \label{fig:prototypeui}
\end{figure*}

The work followed four steps: (i) first, harvesting of datasets and their corresponding metadata from selected data portals. This was followed by (ii) enhancing and preprocessing of the metadata. Preprocessed metadata of the harvested datasets were stored in a PostgreSQL database. We then (iii) developed a prototype application using the Python flask framework (see Figure \ref{fig:prototypeui}). Finally, (iv) the prototype was used to evaluate the suggested strategies both with respect to performance and user-based ratings of relevance. The scripts for datasets/metadata harvesting, and the prototype application can be found on GitHub (\url{https://github.com/brhanebt/recommender}).

\subsection{Data harvesting and preprocessing}
We have harvested machine-readable datasets of GeoJSON formats and corresponding JSON and CSV resources from CKAN-based data portals of three English speaking countries (data.gov.uk\textasciitilde 959, data.gov\textasciitilde 1003, and data.gov.ie\textasciitilde 547 datasets) using the CKAN API. Since the harvested metadata of the datasets didn't always have spatial extent, the spatial metadata enhancement is done (in case of missing spatial extent), either using the metadata or the GeoJSON file's bbox field. The spatial metadata enhancement is done via one of three approaches used in \cite{DeFernandesVasconcelos2017}. First, by collecting minimum and maximum coordinates from each feature in the GeoJSON's feature collection. These are then aggregated into a bounding area of envelopes using OGR Envelope, which returns a tuple (minX, maxX, minY, maxY) \citep{OGRGeometryClassReference}. If geojson features (or feature collections) are missing, the second alternative is applied using DBPedia Spotlight \citep{Mendes2011} to parse place names in title or description in the dataset metadata and OSM Nominatim \citep{OpenStreetMap2018} to find a geometry associated with the place name. If no place name is found in the title or description, the third option uses the data portal's URL and OSM Nominatim to geocode the country's spatial extent as the dataset's spatial extent.

Following the spatial metadata enhancement using either of the above alternatives, the thematic metadata is preprocessed and stored in PostgreSQL. The thematic metadata preprocessing is done by vectorizing (tokenizing) the textual metadata using PostgreSQL's tsvector function for weighted vectorization. Title and tags were given the weight A, and description was ascribed the weight B (see \cite{PostgreSQL2018} for details about weights in PostgreSQL 9). 

\subsection{Experimental setup}
\label{subsec:experimentalsetup}
\subsubsection*{Computer} The prototype was developed, tested and evaluated on a computer with the following characteristics: Computer Manufacturer: HP; Processor: AMD A6-9220 RADEON R4, 5 Compute Core 2c+3G 2.50 GHz;  RAM: 8GB; System type: 64-bit operating system, x64-based processor; Operating System: Windows 10. Any performance value reported later should thus be interpreted in the light of these characteristics. 

\begin{table}[]
    \centering
    \begin{footnotesize}
    \begin{tabular}{|c|c|c|c|c|}
    \hline
& Thematic User Input & Synonym & Hypernym & Hyponym \\\hline
      Weight  & 1.0 & 1.0 & 0.8 & 0.9 \\\hline
    \end{tabular}
    \end{footnotesize}
    \caption{WordNet Query Expansion Weights}
    \label{tab:weightwordnet}
\end{table}

\begin{table}[]
    \centering
    \begin{footnotesize}
    \begin{tabular}{|c|c|c|c|c|}
    \hline
& Thematic User Input & Synonym & IsA & MannerOf \\\hline
Weight  & 1.0 & 1.0 & 0.9 & 0.9 \\\hline
    \end{tabular}
    \end{footnotesize}
    \caption{ConceptNet Query Expansion Weights}
    \label{tab:weightconceptnet}
\end{table}

\subsubsection*{Weights for query expansion} The weights in Table \ref{tab:weightwordnet} and \ref{tab:weightconceptnet}, inspired primarily by previous work \cite{Hsu2008}, were used during the work.

\subsubsection*{Spatial similarity} As mentioned in previous work \cite{Frontiera2008, Cai2011}, two common methods to quantify the spatial similarity of polygons are `area of overlap' (AO) and the Hausdorff distance (HD). Both methods were applied in the study. The spatial similarity was computed in all strategies except the first strategy. 

\subsubsection*{Aggregation} 
The aggregation of the ranking results from both the thematic and spatial ranking is performed using the formula: 
\begin{equation}
Score = N(R(t)) + N(R(s))
\end{equation}
where N(R(t)) is the normalized ranking of the full-text search as returned from PostgreSQL's Ts\_rank. N(R(t)) is obtained by dividing each rank by the range of all scores, to get a value between 0 and 1. N(R(s)) is the normalized ranking of the spatial query result based on either the area of overlap or Hausdorff distance. This too is obtained by dividing a given spatial similarity value by the difference between max and min spatial similarity scores. The normalized area of overlap produces similarity scores ranging from 0 (none similarity) to 1 (complete similarity) while the opposite is true with Hausdorff distance. Therefore, the inverse weighting is used before normalizing the results of Hausdorff distance \citep{Larson2011}. Finally, results are ranked by the aggregated score. This is done for all strategies except the keyword-based baseline strategy that is based only on full-text search.

\subsubsection*{Implementation of strategies} The strategies were implemented as follows: Baseline Keyword (full-text search using both thematic and spatial query terms, and then ranking of results using only PostgreSQL's TS\_Rank); Baseline Spatial (spatially restricted results using the INTERSECTS condition on the metadata stored in PostGIS and the spatial query term; spatial similarity computation using AO/HD; then full-text search on the spatially restricted results using the thematic query term; ranking using the aggregated ranking score above); WordNet QE (expansion of users' input queries with WordNet; then spatially restricted results using the INTERSECTS condition on the metadata stored in PostGIS and the spatial query term; spatial similarity computation using AO/HD; then \textit{weighted} full-text search on the spatially restricted results using the thematic query term and the weights introduced above; ranking using the aggregated ranking score above); ConceptNet QE (same approach as WordNet, except the use of ConceptNet as a source for the expansion instead).



\subsubsection*{Queries} \citet{spink2001searching} reported that the average size of users' web search is 2.4 words. Recent analyses \cite{Koesten2017,Kacprzak2019}, addressing specifically the task of data search, revealed an average query length of 2.44 words and 2.67 respectively. In keeping with these results, the tests were done using about 2 words per query. To evaluate the performance of the strategies (both in terms of response time and number of results), the following four thematic and spatial query combinations were selected after testing multiple ad-hoc queries: 
\begin{enumerate}
    \item Population England (hereafter Q1) 
    \item Learning Wales (hereafter Q2)
    \item Transport Fairfax (hereafter Q4)
    \item Communities Republic of Ireland (hereafter Q4)
\end{enumerate}

\section{Theoretical Evaluation}
\label{sec:theoretical}
Tables \ref{tab:england} to \ref{tab:ireland} presents the results of the performance evaluation. Regularities observed across all four queries are: (i) switching between area of overlap and Hausdorff distance for spatial similarity computation does not have any substantial impact on performance; (ii) query expansion using WordNet synonyms, hypernyms and hyponyms consistently provides the best compromise between diversity (i.e. variety of results) and response time; (iii) query expansion using ConceptNet is, as of now, not recommendable for practical applications ( `for a routine request, the acknowledgment should be within two seconds' \cite{Miller1968}, but results were obtained within a time in the order of 10 seconds);  and (iv) using space as an extra dimension for search boosts results (as evidenced by the differences observed between the `baseline' and `baseline spatial' strategies\footnote{In the baseline strategy, a full-text search of both thematic and spatial keywords is applied on the metadata stored in PostgreSQL without any spatial consideration. On the contrary, the baseline spatial strategy computes first the spatial similarity between the bounding box of the user's toponym and the geographic footprint of the metadata (e.g. using area of overlap or Hausdorff distance). Afterward, the set of results to be returned is restricted, based on the thematic component of the user's query.}). 

\begin{table}
\caption{Performance results for ``Population England''}
\label{tab:england}
\begin{tabular}{|l|l|l|}
\hline
\multicolumn{3}{|c|}{Query Terms: ``Population England'' (Q1)}             \\ \hline
Strategy         & Time (in seconds) & Number of results \\ \hline
Baseline         & 1                 & 3                 \\ \hline
Baseline AO      & 2                 & 21                \\ \hline
Baseline HD      & 2                 & 21                \\ \hline
WordNet 01 AO    & 2                 & 21                \\ \hline
WordNet 01 HD    & 2                 & 21                \\ \hline
WordNet 02 AO    & 4                 & 101               \\ \hline
WordNet 02 HD    & 4                 & 101               \\ \hline
ConceptNet 01 AO & 15                & 21                \\ \hline
ConceptNet 01 HD & 15                & 21                \\ \hline
ConceptNet 02 AO & 15                & 21                \\ \hline
ConceptNet 02 HD & 15                & 21                \\ \hline
\end{tabular}
\end{table}

\begin{table}
\caption{Performance results for ``Learning Wales'' }
\label{tab:wales}
\begin{tabular}{|l|l|l|}
\hline
\multicolumn{3}{|c|}{Query Terms: ``Learning Wales'' (Q2)}        \\ \hline
Strategy         & Time (in seconds) & Number of results \\ \hline
Baseline         & 1                 & 13                \\ \hline
Baseline AO      & 2                 & 14                \\ \hline
Baseline HD      & 2                 & 14                \\ \hline
WordNet 01 AO    & 3                 & 44                \\ \hline
WordNet 01 HD    & 3                 & 44                \\ \hline
WordNet 02 AO    & 4                 & 129               \\ \hline
WordNet 02 HD    & 4                 & 129               \\ \hline
ConceptNet 01 AO & 15                & 31                \\ \hline
ConceptNet 01 HD & 12                & 31                \\ \hline
ConceptNet 02 AO & 15                & 53                \\ \hline
ConceptNet 02 HD & 12                & 53                \\ \hline
\end{tabular}
\end{table}

\begin{table}
\caption{Performance results for ``Transport Fairfax''}
\label{tab:fairfax}
\begin{tabular}{|l|l|l|}
\hline
\multicolumn{3}{|c|}{Query Terms: ``Transport Fairfax'' (Q3)}                 \\ \hline
Strategy         & Time (in seconds) & Number of results \\ \hline
Baseline         & 1                 & 7                \\ \hline
Baseline AO      & 2                 & 16               \\ \hline
Baseline HD      & 2                 & 16               \\ \hline
WordNet 01 AO    & 2                 & 18               \\ \hline
WordNet 01 HD    & 2                 & 18               \\ \hline
WordNet 02 AO    & 2                 & 74               \\ \hline
WordNet 02 HD    & 2                 & 74               \\ \hline
ConceptNet 01 AO & 18                & 21               \\ \hline
ConceptNet 01 HD & 18                & 21               \\ \hline
ConceptNet 02 AO & 18                & 33               \\ \hline
ConceptNet 02 HD & 18                & 33               \\ \hline
\end{tabular}
\end{table}

\begin{table}
\caption{Performance results for ``Communities Republic of Ireland'' }
\label{tab:ireland}
\begin{tabular}{|l|l|l|}
\hline
\multicolumn{3}{|c|}{Query Terms: ``Communities Republic of Ireland'' (Q4)}    \\ \hline
Strategy         & Time (in seconds) & Number of results \\ \hline
Baseline         & 0                 & 0                 \\ \hline
Baseline AO      & 2                 & 79                \\ \hline
Baseline HD      & 2                 & 79                \\ \hline
WordNet 01 AO    & 3                 & 83                \\ \hline
WordNet 01 HD    & 3                 & 83                \\ \hline
WordNet 02 AO    & 5                 & 296               \\ \hline
WordNet 02 HD    & 5                 & 296               \\ \hline
ConceptNet 01 AO & 3                 & 79                \\ \hline
ConceptNet 01 HD & 3                 & 79                \\ \hline
ConceptNet 02 AO & 3                 & 79                \\ \hline
ConceptNet 02 HD & 3                 & 79                \\ \hline
\end{tabular}
\end{table}

\section{User Evaluation}
\label{sec:userstudy}
This section is concerned with the impact of the search strategies on user relevance judgment. As mentioned in Section \ref{subsec:relevance}, relevance in this work is of the form: \textit{rel = (surrogate, request, (topic, task, context))}. Users worked on \textit{surrogate} information resources, that is, representations of datasets consisting of title and description. They expressed their information need using \textit{requests} (i.e. queries in a natural language consisting of two keywords, one thematic and one spatial, see Figure \ref{fig:prototypeui}). The users interacted with four \textit{topics}, which correspond to the four sets of keywords from Section \ref{subsec:experimentalsetup}. The implicit \textit{task} was to `satisfy one's curiosity', and two important elements of \textit{context} here are: the device (the users worked on a desktop computer - instead of a mobile device or a tablet for example); and the user interface design (which provides separate input fields for place names and thematic keywords, see Figure \ref{fig:prototypeui}). Response times for query expansion with ConceptNet exceeded the 10 seconds limit recommended by previous work \cite{Card1991,Miller1968,Nielsen1993}, to keep the user's attention focused on the dialogue with a computer system. For this reason, strategies involving ConceptNet were not included in the user evaluation. Hereafter, s1 refers to Baseline, s1 to Baseline AO, s3 to Baseline HD, s4 to WordNet 01 AO, s5 to WordNet 02 AO, s6 to WordNet 02 AO, and s7 to WordNet 02 AO.

\subsubsection*{Participants and tasks}
16 participants (5 Female) participated in the study. They were recruited via word of mouth at the University of Münster. Most of them (13) had already used open data before the experiment, with reported purposes of usage being: application development (6), scientific hypothesis testing (7), coursework (11) and fun (1). The instruction given to the participants was: 
\begin{quote}
\begin{small}
You are interested in datasets about or related to [TOPIC].\\
(a) Search for these datasets using\\
Query Theme: [KEYWORD1]\\
Query Location: [KEYWORD2]\\
(b) For each of the first seven results, assign a star indicating its relevance to your information need: the more number of stars, the more relevant the result.
\end{small}
\end{quote}

About 2,800 relevance ratings were collected. The same computer was used by all participants for all tasks, and the computer screen was video-recorded. To learn about the rating process itself, participants were asked to fill a short questionnaire after completing all tasks. The study was approved by the institutional ethics board. One participant skipped several query results without rating. Another participant did not rate any query for strategies 1 and 7. Thus, the analysis and results that follow are based on the 14 users who completed the relevance ratings for all seven queries. Queries, where users rated less than the first seven results as requested in the instruction, were also excluded from the analysis. In cases, where a participant provided ratings for more than the first seven results (say 10), only the ratings of the first seven items were considered. In the end, 2219 ratings were included in the analysis.

\subsubsection*{Results}
The discounted cumulative gain (DCG, \cite{Jarvelin2002}) was computed pro query term and user, and then averaged across the 14 users for each query. Q1 returned only three results for strategy s1. The DCG for q1 in s1 was thus computed with rating values of zero for the positions 4, 5, 6, and 7. Table \ref{tab:dcgs} presents the results. To arrive at a conclusion of the `preferred' strategy by the users, a (simple) Borda voting scheme \cite{levin1995introduction} was applied to the results obtained in Table \ref{tab:dcgs}. The seven strategies were ranked according to the mean DCG obtained, and assigned points according to their respective ranks. Table \ref{tab:borda} shows the results. 

\begin{table}
\caption{Average DCGs pro query and strategy}
\label{tab:dcgs}
\begin{tabular}{|r|l|l|l|l|l|}
\hline
 & q1 & q2 & q3 & q4 & \textit{\textbf{Average}} \\ \hline
s1 & 29,34 & 16,15 & 26,8 & \multicolumn{1}{c|}{-} & 24,10 \\ \hline
s2 & 28,56 & 20,61 & 26,17 & 20,84 & 24,05 \\ \hline
s3 & 28,87 & 18,64 & 25,27 & 24,52 & 24,32 \\ \hline
s4 & 28,19 & 21,88 & 24,76 & 20,71 & 23,88 \\ \hline
s5 & 29,61 & 19,57 & 23,99 & 25,77 & 24,7 \\ \hline
s6 & 27,09 & 19,22 & 25,46 & 18,91 & 22,67 \\ \hline
s7 & 22,99 & 18,64 & 20,14 & 24,98 & 21,69 \\ \hline
\textit{\textbf{Average}} & 27,81 & 19,24 & 24,66 & 22,62 & \multicolumn{1}{c|}{-} \\ \hline
\end{tabular}
\end{table}

\begin{table}
\caption{Borda scores for the seven strategies}
\label{tab:borda}
\begin{tabular}{|l|l|l|l|l|c|}
\hline
 & q1 & q2 & q3 & q4 & \multicolumn{1}{l|}{Borda score} \\ \hline
s1 & 5 & 0 & 6 & 0 & 11 \\ \hline
s2 & 3 & 5 & 5 & 3 & 16 \\ \hline
s3 & 4 & 2 & 3 & 4 & 13 \\ \hline
s4 & 2 & 6 & 2 & 2 & 12 \\ \hline
s5 & 6 & 4 & 1 & 6 & 17 \\ \hline
s6 & 1 & 3 & 4 & 1 & 9 \\ \hline
s7 & 0 & 2 & 0 & 5 & 7 \\ \hline
\end{tabular}
\end{table}

\subsubsection*{Feedback about the rating process}
Baseline values for user-based document ratings in the GIR literature are scarce, and so are benchmark values related to the topic. For this reason (and as mentioned above), another objective of the user study was to learn about the rating process itself, to inform similar future studies. Three aspects were examined: the user performance (i.e. in terms of number of results rated pro minute), the variability of the ratings across users, and the perceived difficulty of the rating tasks by users. The analysis of the video recordings showed that users in average rated \textcolor{black}{3.70 items pro minute (standard deviation: 0.9)}. The average coefficient of variation \textbf{of DCGs} across all strategies and queries was 23.6\%, i.e., DCGs of users differed by about 20\% of DCG unit. Table \ref{tab:cvar} shows the coefficients of variations of the DCGs pro query and strategy. 

\begin{table}
\caption{Coefficients of variation (\%) pro query and strategy}
\label{tab:cvar}
\begin{tabular}{|r|l|l|l|l|l|}
\hline
 & q1 (\%) & q2 (\%) & q3 (\%) & q4 (\%) & \textit{\textbf{Average}} \\ \hline
s1 & 11,01 & 21,67 & 16,23 & \multicolumn{1}{c|}{-} & 16,30\% \\ \hline
s2 & 11,41 & 30,71 & 16,01 & 39,59 & 24,43\% \\ \hline
s3 & 11,81 & 32,99 & 14,48 & 29,45 & 22,18\% \\ \hline
s4 & 13,12 & 23,31 & 17,37 & 34,57 & 22,09\% \\ \hline
s5 & 9,29 & 34,70 & 20,47 & 30,77 & 23,81\% \\ \hline
s6 & 10,19 & 41,52 & 15,89 & 44,69 & 28,07\% \\ \hline
s7 & 28,66 & 33,91 & 23,44 & 26,42 & 28,11\% \\ \hline
\textit{\textbf{Average}} & 13,64\% & 31,26\% & 17,69\% & 34,25\% & \multicolumn{1}{c|}{-} \\ \hline
\end{tabular}
\end{table}

\section{Discussion}
\label{sec:discussion}
What is the impact of spatial search strategies on performance and user relevance judgment? Results from the theoretical evaluation suggest that using space as an extra dimension boosts search results. Either area of overlap or Hausdorff distance may be used to do so, if the focus is on response time only. Results from the user evaluation suggest that (i) type of search strategy truly matters, as no two strategies produced a similar DCG for any query (Table \ref{tab:dcgs}); and (ii) more is not always better for query expansion, that is, the combination of synonyms, hypernyms, and hyponyms is detrimental to GIR from the user's point of view (s6 and s7 get lower scores in Table \ref{tab:borda}). Besides, Table \ref{tab:dcgs} suggests that (iii) Hausdorff distance as spatial similarity metric produces \textit{slightly} better results than area of overlap (s6 and s7 left aside) from the user's viewpoint. 

As regards the method, previous work has indicated that a ``challenge for future research in GIR, and more particularly georeferencing, is \textbf{reproducible publishing of methods}, algorithms, datasets, and results such that approaches can be more easily compared across corpora'' (\cite{Purves2018}, emphasis added). The method used in this work has involved two key steps towards a \textit{holistic} assessment of spatial search strategies, namely performance-based and user-based assessment. Each of the steps have had distinct, but complementary goals. The performance-based assessment was useful to find the types of strategies that \textit{can} be used in practical applications; the user-based assessment helped find out the types of strategies that \textit{should} be used in practical applications. Below is a recap of the steps followed at each stage, and pointers to the literature:

\begin{enumerate}

    \item Performance-based assessment (stage 1)
    \begin{itemize}
    \item Queries selection: length between 2 and 3 words \cite{Koesten2017,Kacprzak2019}
    \item Performance metrics selection (e.g. time and number of results)
    \item Selection of strategies for stage 2, based on known cut-off values (e.g. \cite{Card1991,Miller1968,Nielsen1993})
    \end{itemize}
    
    \item User-based assessment (stage 2)
    
    \begin{itemize}
     \item Formal specification of relevance \cite{Mizzaro1998}
     \item Collection of user ratings and analysis based on known metrics (e.g. DCG \cite{Jarvelin2002})
     \item Quantification of user ratings' variability (e.g. using the coefficient of variation)
     \item Vote on the `best strategy' (e.g. using Borda count \cite{levin1995introduction} or other voting schemes) [OPTIONAL]
    \end{itemize}
\end{enumerate}

As such, the method is cognitively grounded, in line with a call from previous work \cite{Purves2018}: ``We suggest that a further challenge for GIR is ... the application of cognitively grounded methods for combination and ranking of spatial and semantically similar information''. There is still the question of the systematic choice of the queries, to which no answer can be given in the current article. Eventually, the queries may be constrained by the application scenario. 

\subsection{Theoretical implications}
The performance-based assessment suggests that more work is needed to make query expansion via ConceptNet scalable and readily available for practical applications. In addition, Table \ref{tab:cvar} illustrates that the queries used as input to the user ratings matter. This points at the need for a typology of spatial `requests' (with request in the sense of \citet{Mizzaro1998}) for GIR. A way of increasing our understanding of what users are looking for could be interviews \textit{after} the rating process, asking \textit{why} they ranked some items the way they did. This greater understanding of the `why' may also help explain the variability of ratings observed during the study. Finally, the results suggest that there may not be a `one-size-fits-all' implementation of autonomous search agents, which assist users in retrieving `relevant' datasets to them. Such agents need to know \textit{why} a user rates relevance in a particular way, and this necessitates an extension of the \citet{Mizzaro1998} model to reflect this.   

\subsection{Practical implications}
A straightforward implication from Table \ref{tab:borda} is that current open data portals may consider query expansion on synonyms (combined with spatial similarity using Hausdorff distance) as an option for their spatial search implementations. The use of spatial search based on the area of overlap seems an equally interesting alternative. Besides, current open data portals return results, probably using one (or none) of the strategies examined in the study. Yet, information about the spatial search strategy used is rarely available. Given, as we have seen, that type of search strategies matter, a documentation of the strategies used would benefit work towards greater transparency in city contexts (besides linking data and visualizing them as suggested by previous work \cite{degbelo2017linked}). Finally, beyond the context of the open government data, the holistic evaluation applied could be a part of a bigger framework to evaluate search strategies of big datasets. In particular, the collection of user-based ratings on sample documents may be a good alternative for cases, where precision/recall might be challenging to obtain due to a large number of data items.

\subsection{Open questions}
Several questions arose during the course of this work, which open up exciting avenues for future research. First, and foremost, GIR research will benefit from \textit{benchmarks} for \textit{all types of relevance judgments} (not just the one examined in this work). These benchmarks should enable us to compare different DCGs for \textit{practical significance}, and different coefficients of variations for \textit{acceptable variability within GIR user studies}. Another question is related to the final decision of the `most appropriate' strategy. The study has brought to light that several parameters influence that decision. The Borda count method was used to reach a conclusion as it reflects broad consensus across participants, but the strategy that appears first or second is still dependent upon the queries included (e.g. Q1, Q2, and Q3 only, or Q1, Q2, Q3, and Q4). Note that this is an inherent limitation of any voting system and `we must always choose between flawed alternatives' \cite{levin1995introduction}. Third, there is the question whether performance values of search strategies and their user-based assessment can be combined into a single metric to facilitate cross-comparison of research results, and whether such a metric is desirable at all. Finally, query reformulation could also be incorporated using the session-based DCG (see \cite{Jarvelin2008}), leading to the open question of the merits of the seven strategies in such a scenario, and of an optimal design of studies to collect a broad range of user ratings while minimizing participants' fatigue.

\subsection{Limitations}
The limited number of queries, and of participants used in the study suggest that the results obtained should be interpreted with caution, and that confidence in their generalizability can only increase as they are replicated. In addition, the results were all computed for a DCG7, that is, the first seven results rated by the users. Computing other DCGs, e.g. DCG3 to DCG6, could help get a more differentiated impression of the relevance feedback given by the users. Last, not least, the experiment did not involve randomization of the appearance of the search strategies, nor the queries. Such randomization would have further minimized the impact of learning effects and possible participants' fatigue.

\section{Conclusion}
\label{sec:conclusion}
This work has investigated the impact of spatial search strategies on performance and user relevance judgment. The search strategies varied according to whether or not they used space as an extra dimension for search, area of overlap or Hausdorff distance as spatial similarity function, and involved query expansion using WordNet or ConceptNet. The performance-based assessment was useful to find the types of strategies that can be used in practical applications; the user-based assessment helped find out the types of strategies that should be used in practical applications. The study revealed that more work is still needed for strategies involving ConceptNet to become readily available for practical applications. In addition, the highest user relevance score was obtained for query expansion on synonyms, combined with spatial similarity computation using Hausdorff distance. Replication of this study, involving more queries and participants, would be beneficial to increase our understanding of observations, which can be generalized to all types of spatial queries and scenarios.

\begin{acks}
The authors gratefully acknowledge funding from the European Commission through the Erasmus Mundus Master in Geospatial Technologies (Erasmus+/Erasmus Mundus program, grant agreement FPA-2012-0191, \url{http://mastergeotech.info/}) and the GEO-C project (H2020-MSCA-ITN-2014, Grant Agreement Number 642332, \url{http://www.geo-c.eu/}).
\end{acks}

\bibliographystyle{ACM-Reference-Format}
\bibliography{gir2019}


\begin{thebibliography}{00}


\ifx \showCODEN    \undefined \def \showCODEN     #1{\unskip}     \fi
\ifx \showDOI      \undefined \def \showDOI       #1{#1}\fi
\ifx \showISBNx    \undefined \def \showISBNx     #1{\unskip}     \fi
\ifx \showISBNxiii \undefined \def \showISBNxiii  #1{\unskip}     \fi
\ifx \showISSN     \undefined \def \showISSN      #1{\unskip}     \fi
\ifx \showLCCN     \undefined \def \showLCCN      #1{\unskip}     \fi
\ifx \shownote     \undefined \def \shownote      #1{#1}          \fi
\ifx \showarticletitle \undefined \def \showarticletitle #1{#1}   \fi
\ifx \showURL      \undefined \def \showURL       {\relax}        \fi
\providecommand\bibfield[2]{#2}
\providecommand\bibinfo[2]{#2}
\providecommand\natexlab[1]{#1}
\providecommand\showeprint[2][]{arXiv:#2}

\bibitem[\protect\citeauthoryear{Ay, Zimmermann, and Kim}{Ay
  et~al\mbox{.}}{2010}]%
        {Ay2010}
\bibfield{author}{\bibinfo{person}{Sakire~Arslan Ay}, \bibinfo{person}{Roger
  Zimmermann}, {and} \bibinfo{person}{Seon~Ho Kim}.}
  \bibinfo{year}{2010}\natexlab{}.
\newblock \showarticletitle{{Relevance ranking in georeferenced video search}}.
\newblock \bibinfo{journal}{{\em Multimedia Systems\/}} \bibinfo{volume}{16},
  \bibinfo{number}{2} (\bibinfo{year}{2010}), \bibinfo{pages}{105--125}.
\newblock
\showISBNx{0942-4962}
\showISSN{09424962}
\showDOI{%
\url{https://doi.org/10.1007/s00530-009-0177-x}}


\bibitem[\protect\citeauthoryear{Azad and Deepak}{Azad and Deepak}{2019}]%
        {Azad2017}
\bibfield{author}{\bibinfo{person}{Hiteshwar~Kumar Azad} {and}
  \bibinfo{person}{Akshay Deepak}.} \bibinfo{year}{2019}\natexlab{}.
\newblock \showarticletitle{Query expansion techniques for information
  retrieval: A survey}.
\newblock \bibinfo{journal}{{\em Information Processing \& Management\/}}
  \bibinfo{volume}{56}, \bibinfo{number}{5} (\bibinfo{year}{2019}),
  \bibinfo{pages}{1698 -- 1735}.
\newblock
\showISSN{0306-4573}
\showDOI{%
\url{https://doi.org/10.1016/j.ipm.2019.05.009}}


\bibitem[\protect\citeauthoryear{Belaid}{Belaid}{2015}]%
        {Belaid2015}
\bibfield{author}{\bibinfo{person}{Rachid Belaid}.}
  \bibinfo{year}{2015}\natexlab{}.
\newblock \bibinfo{title}{{Postgres full-text search is Good Enough!}}
\newblock   (\bibinfo{year}{2015}).
\newblock
\showURL{%
\url{http://blog.lostpropertyhq.com/postgres-full-text-search-is-good-enough/}}


\bibitem[\protect\citeauthoryear{Benitez-Paez, Comber, Trilles, and
  Huerta}{Benitez-Paez et~al\mbox{.}}{2018a}]%
        {Benitez-Paez2018}
\bibfield{author}{\bibinfo{person}{Fernando Benitez-Paez},
  \bibinfo{person}{Alexis Comber}, \bibinfo{person}{Sergio Trilles}, {and}
  \bibinfo{person}{Joaquin Huerta}.} \bibinfo{year}{2018}\natexlab{a}.
\newblock \showarticletitle{{Creating a conceptual framework to improve the
  re-usability of open geographic data in cities}}.
\newblock \bibinfo{journal}{{\em Transactions in GIS\/}} \bibinfo{volume}{22},
  \bibinfo{number}{3} (\bibinfo{date}{jun} \bibinfo{year}{2018}),
  \bibinfo{pages}{806--822}.
\newblock
\showISSN{13611682}
\showDOI{%
\url{https://doi.org/10.1111/tgis.12449}}


\bibitem[\protect\citeauthoryear{Benitez-Paez, Degbelo, Trilles, and
  Huerta}{Benitez-Paez et~al\mbox{.}}{2018b}]%
        {benitez2018roadblock}
\bibfield{author}{\bibinfo{person}{Fernando Benitez-Paez},
  \bibinfo{person}{Auriol Degbelo}, \bibinfo{person}{Sergio Trilles}, {and}
  \bibinfo{person}{Joaquin Huerta}.} \bibinfo{year}{2018}\natexlab{b}.
\newblock \showarticletitle{{Roadblocks hindering the reuse of open geodata in
  Colombia and Spain: A data user's perspective}}.
\newblock \bibinfo{journal}{{\em ISPRS International Journal of
  Geo-Information\/}} \bibinfo{volume}{7}, \bibinfo{number}{1}
  (\bibinfo{date}{dec} \bibinfo{year}{2018}), \bibinfo{pages}{6}.
\newblock
\showISSN{2220-9964}
\showDOI{%
\url{https://doi.org/10.3390/ijgi7010006}}


\bibitem[\protect\citeauthoryear{Bouchoucha, He, and Nie}{Bouchoucha
  et~al\mbox{.}}{2013}]%
        {Bouchoucha2013}
\bibfield{author}{\bibinfo{person}{Arbi Bouchoucha}, \bibinfo{person}{Jing He},
  {and} \bibinfo{person}{Jian-Yun Nie}.} \bibinfo{year}{2013}\natexlab{}.
\newblock \showarticletitle{{Diversified query expansion using conceptnet}}.
\newblock \bibinfo{journal}{{\em Proceedings of the 22nd ACM international
  conference on Conference on information {\&} knowledge management - CIKM
  '13\/}} \bibinfo{number}{October 2013} (\bibinfo{year}{2013}),
  \bibinfo{pages}{1861--1864}.
\newblock
\showISBNx{9781450322638}
\showDOI{%
\url{https://doi.org/10.1145/2505515.2507881}}


\bibitem[\protect\citeauthoryear{Cai}{Cai}{2011}]%
        {Cai2011}
\bibfield{author}{\bibinfo{person}{Guoray Cai}.}
  \bibinfo{year}{2011}\natexlab{}.
\newblock \showarticletitle{{Relevance ranking in Geographical Information
  Retrieval}}.
\newblock \bibinfo{journal}{{\em SIGSPATIAL Special\/}} \bibinfo{volume}{3},
  \bibinfo{number}{2} (\bibinfo{date}{jul} \bibinfo{year}{2011}),
  \bibinfo{pages}{33--36}.
\newblock
\showISSN{19467729}
\showDOI{%
\url{https://doi.org/10.1145/2047296.2047304}}


\bibitem[\protect\citeauthoryear{Card, Robertson, and Mackinlay}{Card
  et~al\mbox{.}}{1991}]%
        {Card1991}
\bibfield{author}{\bibinfo{person}{Stuart~K. Card}, \bibinfo{person}{George~G.
  Robertson}, {and} \bibinfo{person}{Jock~D. Mackinlay}.}
  \bibinfo{year}{1991}\natexlab{}.
\newblock \showarticletitle{{The information visualizer, an information
  workspace}}. In \bibinfo{booktitle}{{\em Proceedings of the SIGCHI conference
  on Human factors in computing systems Reaching through technology - CHI
  '91}}, \bibfield{editor}{\bibinfo{person}{Scott~P. Robertson},
  \bibinfo{person}{Gary~M. Olson}, {and} \bibinfo{person}{Judith~S. Olson}}
  (Eds.). \bibinfo{publisher}{ACM Press}, \bibinfo{address}{New Orleans,
  Louisiana, USA}, \bibinfo{pages}{181--186}.
\newblock
\showISBNx{0897913833}
\showDOI{%
\url{https://doi.org/10.1145/108844.108874}}


\bibitem[\protect\citeauthoryear{Charalabidis, Alexopoulos, and
  Loukis}{Charalabidis et~al\mbox{.}}{2016}]%
        {Charalabidis2016}
\bibfield{author}{\bibinfo{person}{Yannis Charalabidis},
  \bibinfo{person}{Charalampos Alexopoulos}, {and} \bibinfo{person}{Euripidis
  Loukis}.} \bibinfo{year}{2016}\natexlab{}.
\newblock \showarticletitle{{A taxonomy of open government data research areas
  and topics}}.
\newblock \bibinfo{journal}{{\em Journal of Organizational Computing and
  Electronic Commerce\/}} \bibinfo{volume}{26}, \bibinfo{number}{1-2}
  (\bibinfo{date}{apr} \bibinfo{year}{2016}), \bibinfo{pages}{41--63}.
\newblock
\showISSN{1091-9392}
\showDOI{%
\url{https://doi.org/10.1080/10919392.2015.1124720}}


\bibitem[\protect\citeauthoryear{Chen, Song, and Yang}{Chen
  et~al\mbox{.}}{2018}]%
        {Chen2018}
\bibfield{author}{\bibinfo{person}{Zugang Chen}, \bibinfo{person}{Jia Song},
  {and} \bibinfo{person}{Yaping Yang}.} \bibinfo{year}{2018}\natexlab{}.
\newblock \showarticletitle{{Similarity Measurement of Metadata of Geospatial
  Data: An Artificial Neural Network Approach}}.
\newblock \bibinfo{journal}{{\em ISPRS International Journal of
  Geo-Information\/}} \bibinfo{volume}{7}, \bibinfo{number}{3}
  (\bibinfo{year}{2018}).
\newblock
\showDOI{%
\url{https://doi.org/10.3390/ijgi7030090}}


\bibitem[\protect\citeauthoryear{{de Fernandes Vasconcelos}, {de Sousa
  Alencar}, {da Silva Ribeiro}, {Ferreira Rodrigues}, and {de Gomes
  Andrade}}{{de Fernandes Vasconcelos} et~al\mbox{.}}{2017}]%
        {DeFernandesVasconcelos2017}
\bibfield{author}{\bibinfo{person}{Pedro~Arthur {de Fernandes Vasconcelos}},
  \bibinfo{person}{Wensttay {de Sousa Alencar}}, \bibinfo{person}{Victor~Hugo
  {da Silva Ribeiro}}, \bibinfo{person}{Natarajan {Ferreira Rodrigues}}, {and}
  \bibinfo{person}{Fabio {de Gomes Andrade}}.} \bibinfo{year}{2017}\natexlab{}.
\newblock \showarticletitle{{Enabling spatial queries in open government data
  portals}}.
\newblock In \bibinfo{booktitle}{{\em Lecture Notes in Computer Science
  (including subseries Lecture Notes in Artificial Intelligence and Lecture
  Notes in Bioinformatics)}}. Vol.~\bibinfo{volume}{10441 LNCS}.
  \bibinfo{publisher}{Springer, Cham}, \bibinfo{pages}{64--79}.
\newblock
\showISBNx{9783319642475}
\showISSN{16113349}
\showDOI{%
\url{https://doi.org/10.1007/978-3-319-64248-2_6}}


\bibitem[\protect\citeauthoryear{{De Sabbata} and Reichenbacher}{{De Sabbata}
  and Reichenbacher}{2012}]%
        {DeSabbata2012}
\bibfield{author}{\bibinfo{person}{Stefano {De Sabbata}} {and}
  \bibinfo{person}{Tumasch Reichenbacher}.} \bibinfo{year}{2012}\natexlab{}.
\newblock \showarticletitle{{Criteria of geographic relevance: an experimental
  study}}.
\newblock \bibinfo{journal}{{\em International Journal of Geographical
  Information Science\/}} \bibinfo{volume}{26}, \bibinfo{number}{8}
  (\bibinfo{date}{aug} \bibinfo{year}{2012}), \bibinfo{pages}{1495--1520}.
\newblock
\showISSN{1365-8816}
\showDOI{%
\url{https://doi.org/10.1080/13658816.2011.639303}}


\bibitem[\protect\citeauthoryear{Degbelo}{Degbelo}{2017}]%
        {degbelo2017linked}
\bibfield{author}{\bibinfo{person}{Auriol Degbelo}.}
  \bibinfo{year}{2017}\natexlab{}.
\newblock \showarticletitle{{Linked data and visualization: two sides of the
  transparency coin}}. In \bibinfo{booktitle}{{\em Proceedings of the 3rd ACM
  SIGSPATIAL Workshop on Smart Cities and Urban Analytics - UrbanGIS'17}},
  \bibfield{editor}{\bibinfo{person}{Huy~T. Vo} {and} \bibinfo{person}{Bill
  Howe}} (Eds.). \bibinfo{publisher}{ACM Press}, \bibinfo{address}{Los Angeles,
  California, USA}, \bibinfo{pages}{1--8}.
\newblock
\showISBNx{9781450354950}
\showDOI{%
\url{https://doi.org/10.1145/3152178.3152191}}


\bibitem[\protect\citeauthoryear{Degbelo, Granell, Trilles, Bhattacharya, and
  Wissing}{Degbelo et~al\mbox{.}}{2020}]%
        {degbelo2020tell}
\bibfield{author}{\bibinfo{person}{Auriol Degbelo}, \bibinfo{person}{Carlos
  Granell}, \bibinfo{person}{Sergio Trilles}, \bibinfo{person}{Devanjan
  Bhattacharya}, {and} \bibinfo{person}{Jonas Wissing}.}
  \bibinfo{year}{2020}\natexlab{}.
\newblock \showarticletitle{{Tell me how my open Data is re-used: increasing
  transparency through the Open City Toolkit}}.
\newblock In \bibinfo{booktitle}{{\em Open Cities | Open Data}},
  \bibfield{editor}{\bibinfo{person}{Scott Hawken}, \bibinfo{person}{Hoon Han},
  {and} \bibinfo{person}{Chris Pettit}} (Eds.). \bibinfo{publisher}{Palgrave
  Macmillan}, \bibinfo{address}{Singapore}, \bibinfo{pages}{311--330}.
\newblock
\showDOI{%
\url{https://doi.org/10.1007/978-981-13-6605-5_14}}


\bibitem[\protect\citeauthoryear{Degbelo and Kray}{Degbelo and Kray}{2018}]%
        {degbelo2018intelligent}
\bibfield{author}{\bibinfo{person}{Auriol Degbelo} {and}
  \bibinfo{person}{Christian Kray}.} \bibinfo{year}{2018}\natexlab{}.
\newblock \showarticletitle{{Intelligent geovisualizations for open government
  data (vision paper)}}. In \bibinfo{booktitle}{{\em 26th ACM SIGSPATIAL
  International Conference on Advances in Geographic Information Systems}},
  \bibfield{editor}{\bibinfo{person}{Farnoush Banaei-Kashani},
  \bibinfo{person}{Erik~G. Hoel}, \bibinfo{person}{Ralf~Hartmut G{\"{u}}ting},
  \bibinfo{person}{Roberto Tamassia}, {and} \bibinfo{person}{Li~Xiong}} (Eds.).
  \bibinfo{publisher}{ACM Press}, \bibinfo{address}{Seattle, Washington, USA},
  \bibinfo{pages}{77--80}.
\newblock
\showDOI{%
\url{https://doi.org/10.1145/3274895.3274940}}


\bibitem[\protect\citeauthoryear{Degbelo, Trilles, Kray, Bhattacharya,
  Schiestel, Wissing, and Granell}{Degbelo et~al\mbox{.}}{2016}]%
        {degbelo2016designing}
\bibfield{author}{\bibinfo{person}{Auriol Degbelo}, \bibinfo{person}{Sergio
  Trilles}, \bibinfo{person}{Christian Kray}, \bibinfo{person}{Devanjan
  Bhattacharya}, \bibinfo{person}{Nicholas Schiestel}, \bibinfo{person}{Jonas
  Wissing}, {and} \bibinfo{person}{Carlos Granell}.}
  \bibinfo{year}{2016}\natexlab{}.
\newblock \showarticletitle{{Designing semantic application programming
  interfaces for open government data}}.
\newblock \bibinfo{journal}{{\em JeDEM - eJournal of eDemocracy and Open
  Government\/}} \bibinfo{volume}{8}, \bibinfo{number}{2}
  (\bibinfo{year}{2016}), \bibinfo{pages}{21--58}.
\newblock


\bibitem[\protect\citeauthoryear{Ding, Difranzo, Graves, Michaelis, Li,
  Mcguinness, and Hendler}{Ding et~al\mbox{.}}{2010}]%
        {Ding2010}
\bibfield{author}{\bibinfo{person}{Li Ding}, \bibinfo{person}{Dominic
  Difranzo}, \bibinfo{person}{Alvaro Graves}, \bibinfo{person}{James~R
  Michaelis}, \bibinfo{person}{Xian Li}, \bibinfo{person}{Deborah~L
  Mcguinness}, {and} \bibinfo{person}{Jim Hendler}.}
  \bibinfo{year}{2010}\natexlab{}.
\newblock \showarticletitle{{Data-gov Wiki : Towards Linking Government Data}}.
  In \bibinfo{booktitle}{{\em Proceedings of the AAAI 2010 Spring Symposium on
  Linked Data Meets Artificial Intelligence}}.
\newblock


\bibitem[\protect\citeauthoryear{Ding, Lebo, Erickson, Difranzo, Williams, Li,
  Michaelis, Graves, Zheng, Shangguan, Flores, McGuinness, and Hendler}{Ding
  et~al\mbox{.}}{2011}]%
        {Ding2011}
\bibfield{author}{\bibinfo{person}{Li Ding}, \bibinfo{person}{Timothy Lebo},
  \bibinfo{person}{John~S. Erickson}, \bibinfo{person}{Dominic Difranzo},
  \bibinfo{person}{Gregory~Todd Williams}, \bibinfo{person}{Xian Li},
  \bibinfo{person}{James Michaelis}, \bibinfo{person}{Alvaro Graves},
  \bibinfo{person}{Jin~Guang Zheng}, \bibinfo{person}{Zhenning Shangguan},
  \bibinfo{person}{Johanna Flores}, \bibinfo{person}{Deborah~L. McGuinness},
  {and} \bibinfo{person}{James~A. Hendler}.} \bibinfo{year}{2011}\natexlab{}.
\newblock \showarticletitle{{TWC LOGD: A portal for linked open government data
  ecosystems}}.
\newblock \bibinfo{journal}{{\em Journal of Web Semantics\/}}
  \bibinfo{volume}{9}, \bibinfo{number}{3} (\bibinfo{year}{2011}),
  \bibinfo{pages}{325--333}.
\newblock
\showISBNx{1570-8268}
\showISSN{15708268}
\showDOI{%
\url{https://doi.org/10.1016/j.websem.2011.06.002}}


\bibitem[\protect\citeauthoryear{Fritze, Degbelo, Br{\"{u}}ggentisch, and
  Kray}{Fritze et~al\mbox{.}}{2017}]%
        {fritze2017feature}
\bibfield{author}{\bibinfo{person}{Holger Fritze}, \bibinfo{person}{Auriol
  Degbelo}, \bibinfo{person}{Tobias Br{\"{u}}ggentisch}, {and}
  \bibinfo{person}{Christian Kray}.} \bibinfo{year}{2017}\natexlab{}.
\newblock \showarticletitle{{Feature-centric ranking algorithms for
  georeferenced video search}}. In \bibinfo{booktitle}{{\em Proceedings of the
  25th ACM SIGSPATIAL International Conference on Advances in Geographic
  Information Systems (ACM SIGSPATIAL 2017)}},
  \bibfield{editor}{\bibinfo{person}{Erik Hoel}, \bibinfo{person}{Shawn~D.
  Newsam}, \bibinfo{person}{Siva Ravada}, \bibinfo{person}{Roberto Tamassia},
  {and} \bibinfo{person}{Goce Trajcevski}} (Eds.). \bibinfo{publisher}{ACM},
  \bibinfo{address}{Los Angeles, California, USA}, \bibinfo{pages}{1--10}.
\newblock
\showDOI{%
\url{https://doi.org/10.1145/3139958.3139976}}


\bibitem[\protect\citeauthoryear{Frontiera, Larson, and Radke}{Frontiera
  et~al\mbox{.}}{2008}]%
        {Frontiera2008}
\bibfield{author}{\bibinfo{person}{Patricia Frontiera}, \bibinfo{person}{Ray
  Larson}, {and} \bibinfo{person}{John Radke}.}
  \bibinfo{year}{2008}\natexlab{}.
\newblock \showarticletitle{{A comparison of geometric approaches to assessing
  spatial similarity for GIR}}.
\newblock \bibinfo{journal}{{\em International Journal of Geographical
  Information Science\/}} \bibinfo{volume}{22}, \bibinfo{number}{3}
  (\bibinfo{date}{mar} \bibinfo{year}{2008}), \bibinfo{pages}{337--360}.
\newblock
\showISSN{1365-8816}
\showDOI{%
\url{https://doi.org/10.1080/13658810701626293}}


\bibitem[\protect\citeauthoryear{{GDAL/OGR contributors}}{{GDAL/OGR
  contributors}}{2018}]%
        {OGRGeometryClassReference}
\bibfield{author}{\bibinfo{person}{{GDAL/OGR contributors}}.}
  \bibinfo{year}{2018}\natexlab{}.
\newblock \bibinfo{title}{{GDAL/OGR} Geospatial Data Abstraction software
  Library}.
\newblock   (\bibinfo{year}{2018}).
\newblock
\showURL{%
\url{https://www.gdal.org/classOGRGeometry.html}}


\bibitem[\protect\citeauthoryear{Hahmann and Burghardt}{Hahmann and
  Burghardt}{2013}]%
        {Hahmann2013}
\bibfield{author}{\bibinfo{person}{Stefan Hahmann} {and} \bibinfo{person}{Dirk
  Burghardt}.} \bibinfo{year}{2013}\natexlab{}.
\newblock \showarticletitle{{How much information is geospatially referenced?
  Networks and cognition}}.
\newblock \bibinfo{journal}{{\em International Journal of Geographical
  Information Science\/}} \bibinfo{volume}{27}, \bibinfo{number}{6}
  (\bibinfo{year}{2013}), \bibinfo{pages}{1171--1189}.
\newblock
\showISBNx{1365-8816}
\showISSN{1365-8816}
\showDOI{%
\url{https://doi.org/10.1080/13658816.2012.743664}}


\bibitem[\protect\citeauthoryear{Hinz and Bill}{Hinz and Bill}{2018}]%
        {Hinz2018a}
\bibfield{author}{\bibinfo{person}{Matthias Hinz} {and} \bibinfo{person}{Ralf
  Bill}.} \bibinfo{year}{2018}\natexlab{}.
\newblock \showarticletitle{{Mapping the landscape of open geodata}}. In
  \bibinfo{booktitle}{{\em Geospatial Technologies for All : short papers,
  posters and poster abstracts of the 21th AGILE Conference on Geographic
  Information Science (AGILE 2018)}}, \bibfield{editor}{\bibinfo{person}{Ali
  Mansourian}, \bibinfo{person}{Petter Pilesj{\"{o}}}, \bibinfo{person}{Lars
  Harrie}, {and} \bibinfo{person}{Ron van Lammeren}} (Eds.).
  \bibinfo{publisher}{Online Proceedings}, \bibinfo{address}{Lund, Sweden}.
\newblock


\bibitem[\protect\citeauthoryear{Hsu, Tsai, and Chen}{Hsu
  et~al\mbox{.}}{2006}]%
        {Hsu2006}
\bibfield{author}{\bibinfo{person}{Ming-Hung Hsu}, \bibinfo{person}{Ming-Feng
  Tsai}, {and} \bibinfo{person}{Hsin-Hsi Chen}.}
  \bibinfo{year}{2006}\natexlab{}.
\newblock \showarticletitle{Query Expansion with ConceptNet and WordNet: An
  Intrinsic Comparison}. In \bibinfo{booktitle}{{\em Information Retrieval
  Technology (AIRS 2006)}}. \bibinfo{publisher}{Springer Berlin Heidelberg},
  \bibinfo{address}{Berlin, Heidelberg}, \bibinfo{pages}{1--13}.
\newblock
\showDOI{%
\url{https://doi.org/10.1007/11880592_1}}


\bibitem[\protect\citeauthoryear{Hsu, Tsai, and Chen}{Hsu
  et~al\mbox{.}}{2008}]%
        {Hsu2008}
\bibfield{author}{\bibinfo{person}{Ming-Hung Hsu}, \bibinfo{person}{Ming-Feng
  Tsai}, {and} \bibinfo{person}{Hsin-Hsi Chen}.}
  \bibinfo{year}{2008}\natexlab{}.
\newblock \showarticletitle{{Combining WordNet and ConceptNet for automatic
  query expansion: A learning approach}}. In \bibinfo{booktitle}{{\em 4th Asia
  Infomation Retrieval Symposium (AIRS 2008)}},
  \bibfield{editor}{\bibinfo{person}{Hang Li}, \bibinfo{person}{Ting Liu},
  \bibinfo{person}{Wei-Ying Ma}, \bibinfo{person}{Tetsuya Sakai},
  \bibinfo{person}{Kam-Fai Wong}, {and} \bibinfo{person}{Guodong Zhou}} (Eds.).
  \bibinfo{address}{Harbin, China}, \bibinfo{pages}{213--224}.
\newblock
\showDOI{%
\url{https://doi.org/10.1007/978-3-540-68636-1_21}}


\bibitem[\protect\citeauthoryear{J{\"{a}}rvelin and
  Kek{\"{a}}l{\"{a}}inen}{J{\"{a}}rvelin and Kek{\"{a}}l{\"{a}}inen}{2002}]%
        {Jarvelin2002}
\bibfield{author}{\bibinfo{person}{Kalervo J{\"{a}}rvelin} {and}
  \bibinfo{person}{Jaana Kek{\"{a}}l{\"{a}}inen}.}
  \bibinfo{year}{2002}\natexlab{}.
\newblock \showarticletitle{{Cumulated gain-based evaluation of IR
  techniques}}.
\newblock \bibinfo{journal}{{\em ACM Transactions on Information Systems\/}}
  \bibinfo{volume}{20}, \bibinfo{number}{4} (\bibinfo{year}{2002}),
  \bibinfo{pages}{422--446}.
\newblock
\showISSN{10468188}
\showDOI{%
\url{https://doi.org/10.1145/582415.582418}}


\bibitem[\protect\citeauthoryear{J{\"{a}}rvelin, Price, Delcambre, and
  Nielsen}{J{\"{a}}rvelin et~al\mbox{.}}{2008}]%
        {Jarvelin2008}
\bibfield{author}{\bibinfo{person}{Kalervo J{\"{a}}rvelin},
  \bibinfo{person}{Susan~L. Price}, \bibinfo{person}{Lois M.~L. Delcambre},
  {and} \bibinfo{person}{Marianne~Lykke Nielsen}.}
  \bibinfo{year}{2008}\natexlab{}.
\newblock \showarticletitle{{Discounted cumulated gain based evaluation of
  multiple-query IR sessions}}. In \bibinfo{booktitle}{{\em Proceedings of the
  30th European Conference on Information Retrieval}},
  \bibfield{editor}{\bibinfo{person}{C.~Macdonald}, \bibinfo{person}{I.~Ounis},
  \bibinfo{person}{V.~Plachouras}, \bibinfo{person}{I.~Ruthven}, {and}
  \bibinfo{person}{R.W. White}} (Eds.). \bibinfo{address}{Glasgow, UK},
  \bibinfo{pages}{4--15}.
\newblock
\showDOI{%
\url{https://doi.org/10.1007/978-3-540-78646-7_4}}


\bibitem[\protect\citeauthoryear{Jiang, Li, Yang, Hu, Armstrong, Huang, Moroni,
  McGibbney, Greguska, and Finch}{Jiang et~al\mbox{.}}{2018}]%
        {Jiang2018}
\bibfield{author}{\bibinfo{person}{Yongyao Jiang}, \bibinfo{person}{Yun Li},
  \bibinfo{person}{Chaowei Yang}, \bibinfo{person}{Fei Hu},
  \bibinfo{person}{Edward~M Armstrong}, \bibinfo{person}{Thomas Huang},
  \bibinfo{person}{David Moroni}, \bibinfo{person}{Lewis~J McGibbney},
  \bibinfo{person}{Frank Greguska}, {and} \bibinfo{person}{Christopher~J
  Finch}.} \bibinfo{year}{2018}\natexlab{}.
\newblock \showarticletitle{{A Smart Web-Based Geospatial Data Discovery System
  with Oceanographic Data as an Example}}.
\newblock \bibinfo{journal}{{\em ISPRS International Journal of
  Geo-Information\/}}  \bibinfo{volume}{7} (\bibinfo{year}{2018}).
\newblock
Issue 2.
\showDOI{%
\url{https://doi.org/10.3390/ijgi7020062}}


\bibitem[\protect\citeauthoryear{Johnson, Sieber, Scassa, Stephens, and
  Robinson}{Johnson et~al\mbox{.}}{2017}]%
        {Johnson2017}
\bibfield{author}{\bibinfo{person}{Peter~A. Johnson}, \bibinfo{person}{Renee
  Sieber}, \bibinfo{person}{Teresa Scassa}, \bibinfo{person}{Monica Stephens},
  {and} \bibinfo{person}{Pamela Robinson}.} \bibinfo{year}{2017}\natexlab{}.
\newblock \showarticletitle{{The cost(s) of geospatial open data}}.
\newblock \bibinfo{journal}{{\em Transactions in GIS\/}} \bibinfo{volume}{21},
  \bibinfo{number}{3} (\bibinfo{date}{jun} \bibinfo{year}{2017}),
  \bibinfo{pages}{434--445}.
\newblock
\showISSN{13611682}
\showDOI{%
\url{https://doi.org/10.1111/tgis.12283}}


\bibitem[\protect\citeauthoryear{Kacprzak, Koesten, Ib{\'{a}}{\~{n}}ez, Blount,
  Tennison, and Simperl}{Kacprzak et~al\mbox{.}}{2019}]%
        {Kacprzak2019}
\bibfield{author}{\bibinfo{person}{Emilia Kacprzak}, \bibinfo{person}{Laura
  Koesten}, \bibinfo{person}{Luis-Daniel Ib{\'{a}}{\~{n}}ez},
  \bibinfo{person}{Tom Blount}, \bibinfo{person}{Jeni Tennison}, {and}
  \bibinfo{person}{Elena Simperl}.} \bibinfo{year}{2019}\natexlab{}.
\newblock \showarticletitle{{Characterising dataset search - An analysis of
  search logs and data requests}}.
\newblock \bibinfo{journal}{{\em Journal of Web Semantics\/}}
  \bibinfo{volume}{55} (\bibinfo{date}{mar} \bibinfo{year}{2019}),
  \bibinfo{pages}{37--55}.
\newblock
\showISSN{15708268}
\showDOI{%
\url{https://doi.org/10.1016/j.websem.2018.11.003}}


\bibitem[\protect\citeauthoryear{Kilic and Aksakalli}{Kilic and
  Aksakalli}{2016}]%
        {Klc2016}
\bibfield{author}{\bibinfo{person}{Gur Kilic} {and}
  \bibinfo{person}{Karabey~sil Aksakalli}.} \bibinfo{year}{2016}\natexlab{}.
\newblock \showarticletitle{{Comparison of Solr and Elasticsearch Among Popular
  Full Text Search Engines and Their Security Analysis}}.
\newblock  (\bibinfo{year}{2016}).
\newblock
\showISBNx{9781509016792}
\showDOI{%
\url{https://doi.org/10.13140/RG.2.2.24563.32803}}


\bibitem[\protect\citeauthoryear{Koesten, Kacprzak, Tennison, and
  Simperl}{Koesten et~al\mbox{.}}{2017}]%
        {Koesten2017}
\bibfield{author}{\bibinfo{person}{Laura~M. Koesten}, \bibinfo{person}{Emilia
  Kacprzak}, \bibinfo{person}{Jenifer F.~A. Tennison}, {and}
  \bibinfo{person}{Elena Simperl}.} \bibinfo{year}{2017}\natexlab{}.
\newblock \showarticletitle{{The trials and tribulations of working with
  structured data}}. In \bibinfo{booktitle}{{\em Proceedings of the 2017 CHI
  Conference on Human Factors in Computing Systems - CHI '17}},
  \bibfield{editor}{\bibinfo{person}{Gloria Mark}, \bibinfo{person}{Susan~R.
  Fussell}, \bibinfo{person}{Cliff Lampe}, \bibinfo{person}{m.~c. Schraefel},
  \bibinfo{person}{Juan~Pablo Hourcade}, \bibinfo{person}{Caroline Appert},
  {and} \bibinfo{person}{Daniel Wigdor}} (Eds.). \bibinfo{publisher}{ACM
  Press}, \bibinfo{address}{Denver, Colorado, USA},
  \bibinfo{pages}{1277--1289}.
\newblock
\showISBNx{9781450346559}
\showDOI{%
\url{https://doi.org/10.1145/3025453.3025838}}


\bibitem[\protect\citeauthoryear{Kuo and Chou}{Kuo and Chou}{2019}]%
        {kuo2019metadata}
\bibfield{author}{\bibinfo{person}{Chiao-Ling Kuo} {and}
  \bibinfo{person}{Han-Chuan Chou}.} \bibinfo{year}{2019}\natexlab{}.
\newblock \showarticletitle{{Metadata assessment for efficient open data
  retrieval}}. In \bibinfo{booktitle}{{\em Accepted Short Papers and Posters
  from the 22nd AGILE Conference on Geo-information Science (AGILE 2019)}},
  \bibfield{editor}{\bibinfo{person}{Phaedon Kyriakidis},
  \bibinfo{person}{Diofantos Hadjimitsis}, \bibinfo{person}{Dimitrios
  Skarlatos}, {and} \bibinfo{person}{Ali Mansourian}} (Eds.).
  \bibinfo{address}{Cyprus, Greece}.
\newblock


\bibitem[\protect\citeauthoryear{Lacasta, Lopez-Pellicer, Espejo-Garc{\'{i}}a,
  Nogueras-Iso, and Zarazaga-Soria}{Lacasta et~al\mbox{.}}{2017}]%
        {Lacasta2017}
\bibfield{author}{\bibinfo{person}{Javier Lacasta}, \bibinfo{person}{F.~Javier
  Lopez-Pellicer}, \bibinfo{person}{Borja Espejo-Garc{\'{i}}a},
  \bibinfo{person}{Javier Nogueras-Iso}, {and} \bibinfo{person}{F.~Javier
  Zarazaga-Soria}.} \bibinfo{year}{2017}\natexlab{}.
\newblock \showarticletitle{{Aggregation-based information retrieval system for
  geospatial data catalogs}}.
\newblock \bibinfo{journal}{{\em International Journal of Geographical
  Information Science\/}} \bibinfo{volume}{31}, \bibinfo{number}{8}
  (\bibinfo{year}{2017}), \bibinfo{pages}{1583--1605}.
\newblock
\showISSN{13623087}
\showDOI{%
\url{https://doi.org/10.1080/13658816.2017.1319949}}


\bibitem[\protect\citeauthoryear{Lafia, Turner, and Kuhn}{Lafia
  et~al\mbox{.}}{2018}]%
        {lafia2018improving}
\bibfield{author}{\bibinfo{person}{Sara Lafia}, \bibinfo{person}{Andrew
  Turner}, {and} \bibinfo{person}{Werner Kuhn}.}
  \bibinfo{year}{2018}\natexlab{}.
\newblock \showarticletitle{{Improving discovery of open civic data}}. In
  \bibinfo{booktitle}{{\em 10th International Conference on Geographic
  Information Science (GIScience 2018)}},
  \bibfield{editor}{\bibinfo{person}{Stephan Winter}, \bibinfo{person}{Amy
  Griffin}, {and} \bibinfo{person}{Monika Sester}} (Eds.),
  Vol.~\bibinfo{volume}{114}. \bibinfo{publisher}{Schloss
  Dagstuhl-Leibniz-Zentrum fuer Informatik}, \bibinfo{address}{Melbourne,
  Australia}, \bibinfo{pages}{9:1--9:15}.
\newblock
\showISBNx{978-3-95977-083-5}
\showISSN{1868-8969}
\showDOI{%
\url{https://doi.org/10.4230/LIPIcs.GISCIENCE.2018.9}}


\bibitem[\protect\citeauthoryear{Larson}{Larson}{2011}]%
        {Larson2011}
\bibfield{author}{\bibinfo{person}{Ray~R. Larson}.}
  \bibinfo{year}{2011}\natexlab{}.
\newblock \showarticletitle{Ranking Approaches for GIR}.
\newblock \bibinfo{journal}{{\em SIGSPATIAL Special\/}} \bibinfo{volume}{3},
  \bibinfo{number}{2} (\bibinfo{date}{July} \bibinfo{year}{2011}),
  \bibinfo{pages}{37--41}.
\newblock
\showISSN{1946-7729}
\showDOI{%
\url{https://doi.org/10.1145/2047296.2047305}}


\bibitem[\protect\citeauthoryear{Levin and Nalebuff}{Levin and
  Nalebuff}{1995}]%
        {levin1995introduction}
\bibfield{author}{\bibinfo{person}{Jonathan Levin} {and} \bibinfo{person}{Barry
  Nalebuff}.} \bibinfo{year}{1995}\natexlab{}.
\newblock \showarticletitle{{An introduction to vote-counting schemes}}.
\newblock \bibinfo{journal}{{\em The Journal of Economic Perspectives\/}}
  \bibinfo{volume}{9}, \bibinfo{number}{1} (\bibinfo{year}{1995}),
  \bibinfo{pages}{3--26}.
\newblock


\bibitem[\protect\citeauthoryear{Mendes, Jakob, Garc\'{\i}a-Silva, and
  Bizer}{Mendes et~al\mbox{.}}{2011}]%
        {Mendes2011}
\bibfield{author}{\bibinfo{person}{Pablo~N. Mendes}, \bibinfo{person}{Max
  Jakob}, \bibinfo{person}{Andr{\'e}s Garc\'{\i}a-Silva}, {and}
  \bibinfo{person}{Christian Bizer}.} \bibinfo{year}{2011}\natexlab{}.
\newblock \showarticletitle{DBpedia Spotlight: Shedding Light on the Web of
  Documents}. In \bibinfo{booktitle}{{\em Proceedings of the 7th International
  Conference on Semantic Systems (I-Semantics '11)}}. \bibinfo{publisher}{ACM},
  \bibinfo{address}{Graz, Austria}, \bibinfo{pages}{1--8}.
\newblock
\showISBNx{978-1-4503-0621-8}
\showDOI{%
\url{https://doi.org/10.1145/2063518.2063519}}


\bibitem[\protect\citeauthoryear{Miller, Beckwith, Fellbaum, Gross, and
  Miller}{Miller et~al\mbox{.}}{1990}]%
        {Miller1990}
\bibfield{author}{\bibinfo{person}{George~A. Miller}, \bibinfo{person}{Richard
  Beckwith}, \bibinfo{person}{Christiane Fellbaum}, \bibinfo{person}{Derek
  Gross}, {and} \bibinfo{person}{Katherine~J. Miller}.}
  \bibinfo{year}{1990}\natexlab{}.
\newblock \showarticletitle{{Introduction to WordNet: An On-line Lexical
  Database*}}.
\newblock \bibinfo{journal}{{\em International Journal of Lexicography\/}}
  \bibinfo{volume}{3}, \bibinfo{number}{4} (\bibinfo{date}{12}
  \bibinfo{year}{1990}), \bibinfo{pages}{235--244}.
\newblock
\showISSN{0950-3846}
\showDOI{%
\url{https://doi.org/10.1093/ijl/3.4.235}}


\bibitem[\protect\citeauthoryear{Miller}{Miller}{1968}]%
        {Miller1968}
\bibfield{author}{\bibinfo{person}{Robert~B. Miller}.}
  \bibinfo{year}{1968}\natexlab{}.
\newblock \showarticletitle{{Response time in man-computer conversational
  transactions}}. In \bibinfo{booktitle}{{\em AFIPS'68 (Fall, part I) -
  Proceedings of the December 9-11, 1968, fall joint computer conference, part
  I}}. \bibinfo{publisher}{ACM Press}, \bibinfo{address}{San Francisco,
  California, USA}, \bibinfo{pages}{267--277}.
\newblock
\showDOI{%
\url{https://doi.org/10.1145/1476589.1476628}}


\bibitem[\protect\citeauthoryear{{Millette} and {Hosein}}{{Millette} and
  {Hosein}}{2016}]%
        {7460350}
\bibfield{author}{\bibinfo{person}{C. {Millette}} {and} \bibinfo{person}{P.
  {Hosein}}.} \bibinfo{year}{2016}\natexlab{}.
\newblock \showarticletitle{A consumer focused open data platform}. In
  \bibinfo{booktitle}{{\em 2016 3rd MEC International Conference on Big Data
  and Smart City (ICBDSC)}}. \bibinfo{pages}{1--6}.
\newblock
\showDOI{%
\url{https://doi.org/10.1109/ICBDSC.2016.7460350}}


\bibitem[\protect\citeauthoryear{MIT}{MIT}{2019}]%
        {MIT2019}
\bibfield{author}{\bibinfo{person}{Open Mind Common~Sense MIT}.}
  \bibinfo{year}{2019}\natexlab{}.
\newblock \bibinfo{title}{{ConceptNet}}.
\newblock   (\bibinfo{year}{2019}).
\newblock
\showURL{%
\url{http://conceptnet.io/}}


\bibitem[\protect\citeauthoryear{MIT-Media-Lab}{MIT-Media-Lab}{2018}]%
        {Relation49:online}
\bibfield{author}{\bibinfo{person}{MIT-Media-Lab}.}
  \bibinfo{year}{2018}\natexlab{}.
\newblock \bibinfo{title}{Relations in ConceptNet 5}.
\newblock
  \bibinfo{howpublished}{\url{https://github.com/commonsense/conceptnet5/wiki/Relations}}.
    (\bibinfo{date}{Apr} \bibinfo{year}{2018}).
\newblock
\newblock
\shownote{(Accessed on 02/22/2019).}


\bibitem[\protect\citeauthoryear{Mizzaro}{Mizzaro}{1998}]%
        {Mizzaro1998}
\bibfield{author}{\bibinfo{person}{Stefano Mizzaro}.}
  \bibinfo{year}{1998}\natexlab{}.
\newblock \showarticletitle{{How many relevances in information retrieval?}}
\newblock \bibinfo{journal}{{\em Interacting with Computers\/}}
  \bibinfo{volume}{10}, \bibinfo{number}{3} (\bibinfo{date}{jun}
  \bibinfo{year}{1998}), \bibinfo{pages}{303--320}.
\newblock
\showISSN{09535438}
\showDOI{%
\url{https://doi.org/10.1016/S0953-5438(98)00012-5}}


\bibitem[\protect\citeauthoryear{Neumaier and Polleres}{Neumaier and
  Polleres}{2019}]%
        {Neumaier2019}
\bibfield{author}{\bibinfo{person}{Sebastian Neumaier} {and}
  \bibinfo{person}{Axel Polleres}.} \bibinfo{year}{2019}\natexlab{}.
\newblock \showarticletitle{{Enabling spatio-temporal search in open data}}.
\newblock \bibinfo{journal}{{\em Journal of Web Semantics\/}}
  \bibinfo{volume}{55} (\bibinfo{date}{mar} \bibinfo{year}{2019}),
  \bibinfo{pages}{21--36}.
\newblock
\showISSN{15708268}
\showDOI{%
\url{https://doi.org/10.1016/j.websem.2018.12.007}}


\bibitem[\protect\citeauthoryear{Nielsen}{Nielsen}{1993}]%
        {Nielsen1993}
\bibfield{author}{\bibinfo{person}{Jakob Nielsen}.}
  \bibinfo{year}{1993}\natexlab{}.
\newblock \bibinfo{title}{{Response times: the 3 important limits}}.
\newblock   (\bibinfo{year}{1993}).
\newblock
\showURL{%
\url{https://www.nngroup.com/articles/response-times-3-important-limits/}}


\bibitem[\protect\citeauthoryear{Noy, Burgess, and Brickley}{Noy
  et~al\mbox{.}}{2019}]%
        {Noy2019}
\bibfield{author}{\bibinfo{person}{Natasha Noy}, \bibinfo{person}{Matthew
  Burgess}, {and} \bibinfo{person}{Dan Brickley}.}
  \bibinfo{year}{2019}\natexlab{}.
\newblock \showarticletitle{{Google Dataset Search: Building a search engine
  for datasets in an open Web ecosystem}}. In \bibinfo{booktitle}{{\em
  Proceedings of The Web Conference (WebConf'2019)}},
  \bibfield{editor}{\bibinfo{person}{Ling Liu}, \bibinfo{person}{Ryen~W.
  White}, \bibinfo{person}{Amin Mantrach}, \bibinfo{person}{Fabrizio
  Silvestri}, \bibinfo{person}{Julian~J. McAuley}, \bibinfo{person}{Ricardo~S.
  Baeza-Yates}, {and} \bibinfo{person}{Leila Zia}} (Eds.).
  \bibinfo{publisher}{ACM Press}, \bibinfo{address}{San Francisco, California,
  USA}, \bibinfo{pages}{1365--1375}.
\newblock
\showISBNx{9781450366748}
\showDOI{%
\url{https://doi.org/10.1145/3308558.3313685}}


\bibitem[\protect\citeauthoryear{{Open Knowledge Foundation}}{{Open Knowledge
  Foundation}}{2009}]%
        {OpenKnowledgeFoundation2009}
\bibfield{author}{\bibinfo{person}{{Open Knowledge Foundation}}.}
  \bibinfo{year}{2009}\natexlab{}.
\newblock \bibinfo{title}{{User guide — CKAN 2.7.3 documentation}}.
\newblock   (\bibinfo{year}{2009}).
\newblock
\showURL{%
\url{https://docs.ckan.org/en/ckan-2.7.3/user-guide.html}}


\bibitem[\protect\citeauthoryear{{Open Street Map}}{{Open Street Map}}{2018}]%
        {OpenStreetMap2018}
\bibfield{author}{\bibinfo{person}{{Open Street Map}}.}
  \bibinfo{year}{2018}\natexlab{}.
\newblock \bibinfo{title}{{Nominatim - OpenStreetMap Wiki}}.
\newblock   (\bibinfo{year}{2018}).
\newblock
\showURL{%
\url{https://wiki.openstreetmap.org/wiki/Nominatim}}


\bibitem[\protect\citeauthoryear{Pal, Mitra, and Datta}{Pal
  et~al\mbox{.}}{2014}]%
        {Pal2013}
\bibfield{author}{\bibinfo{person}{Dipasree Pal}, \bibinfo{person}{Mandar
  Mitra}, {and} \bibinfo{person}{Kalyankumar Datta}.}
  \bibinfo{year}{2014}\natexlab{}.
\newblock \showarticletitle{Improving Query Expansion Using WordNet}.
\newblock \bibinfo{journal}{{\em J. Assoc. Inf. Sci. Technol.\/}}
  \bibinfo{volume}{65}, \bibinfo{number}{12} (\bibinfo{date}{Dec.}
  \bibinfo{year}{2014}), \bibinfo{pages}{2469--2478}.
\newblock
\showISSN{2330-1635}
\showDOI{%
\url{https://doi.org/10.1002/asi.23143}}


\bibitem[\protect\citeauthoryear{PostgreSQL}{PostgreSQL}{2018}]%
        {PostgreSQL2018}
\bibfield{author}{\bibinfo{person}{PostgreSQL}.}
  \bibinfo{year}{2018}\natexlab{}.
\newblock \bibinfo{title}{{PostgreSQL: Documentation: 9.0: Controlling Text
  Search}}.
\newblock   (\bibinfo{year}{2018}).
\newblock
\showURL{%
\url{https://www.postgresql.org/docs/9.0/textsearch-controls.html}}


\bibitem[\protect\citeauthoryear{{PostgreSQL Global Development
  Group}}{{PostgreSQL Global Development Group}}{2016}]%
        {PostgreSQLGlobalDevelopmentGroup2016}
\bibfield{author}{\bibinfo{person}{{PostgreSQL Global Development Group}}.}
  \bibinfo{year}{2016}\natexlab{}.
\newblock \bibinfo{title}{{PostgreSQL Full Text Search}}.
\newblock   (\bibinfo{year}{2016}).
\newblock
\showURL{%
\url{http://www.postgresql.org/docs/8.3/static/textsearch-intro.html}}


\bibitem[\protect\citeauthoryear{Purves, Clough, Jones, Hall, and
  Murdock}{Purves et~al\mbox{.}}{2018}]%
        {Purves2018}
\bibfield{author}{\bibinfo{person}{Ross~S. Purves}, \bibinfo{person}{Paul
  Clough}, \bibinfo{person}{Christopher~B. Jones}, \bibinfo{person}{Mark~H.
  Hall}, {and} \bibinfo{person}{Vanessa Murdock}.}
  \bibinfo{year}{2018}\natexlab{}.
\newblock \showarticletitle{{Geographic information retrieval: progress and
  challenges in spatial search of text}}.
\newblock \bibinfo{journal}{{\em Foundations and Trends{\textregistered} in
  Information Retrieval\/}} \bibinfo{volume}{12}, \bibinfo{number}{2-3}
  (\bibinfo{year}{2018}), \bibinfo{pages}{164--318}.
\newblock
\showISSN{1554-0669}
\showDOI{%
\url{https://doi.org/10.1561/1500000034}}


\bibitem[\protect\citeauthoryear{Reichenbacher, {De Sabbata}, Purves, and
  Fabrikant}{Reichenbacher et~al\mbox{.}}{2016}]%
        {Reichenbacher2016}
\bibfield{author}{\bibinfo{person}{Tumasch Reichenbacher},
  \bibinfo{person}{Stefano {De Sabbata}}, \bibinfo{person}{Ross~S. Purves},
  {and} \bibinfo{person}{Sara~I. Fabrikant}.} \bibinfo{year}{2016}\natexlab{}.
\newblock \showarticletitle{{Assessing geographic relevance for mobile search:
  A computational model and its validation via crowdsourcing}}.
\newblock \bibinfo{journal}{{\em Journal of the Association for Information
  Science and Technology\/}} \bibinfo{volume}{67}, \bibinfo{number}{11}
  (\bibinfo{date}{nov} \bibinfo{year}{2016}), \bibinfo{pages}{2620--2634}.
\newblock
\showISSN{23301635}
\showDOI{%
\url{https://doi.org/10.1002/asi.23625}}


\bibitem[\protect\citeauthoryear{Rivas, Iglesias, and Borrajo}{Rivas
  et~al\mbox{.}}{2014}]%
        {Rivas2014}
\bibfield{author}{\bibinfo{person}{A.~R. Rivas}, \bibinfo{person}{E.~L.
  Iglesias}, {and} \bibinfo{person}{L. Borrajo}.}
  \bibinfo{year}{2014}\natexlab{}.
\newblock \showarticletitle{{Study of query expansion techniques and their
  application in the biomedical information retrieval}}.
\newblock \bibinfo{journal}{{\em The Scientific World Journal\/}}
  \bibinfo{volume}{2014} (\bibinfo{year}{2014}).
\newblock
\showISBNx{1537-744x}
\showISSN{1537744X}
\showDOI{%
\url{https://doi.org/10.1155/2014/132158}}


\bibitem[\protect\citeauthoryear{Rozell, Erickson, and Hendler}{Rozell
  et~al\mbox{.}}{2012}]%
        {Rozell2012}
\bibfield{author}{\bibinfo{person}{E Rozell}, \bibinfo{person}{J Erickson},
  {and} \bibinfo{person}{J Hendler}.} \bibinfo{year}{2012}\natexlab{}.
\newblock \showarticletitle{{From international open government dataset search
  to discovery: A semantic web service approach}}.
\newblock \bibinfo{journal}{{\em ACM International Conference Proceeding
  Series\/}} (\bibinfo{year}{2012}), \bibinfo{pages}{480--481}.
\newblock
\showISBNx{9781450312004}
\showDOI{%
\url{https://doi.org/10.1145/2463728.2463827}}


\bibitem[\protect\citeauthoryear{Speer and Havasi}{Speer and Havasi}{2012}]%
        {Speer2012RepresentingGR}
\bibfield{author}{\bibinfo{person}{Robyn Speer} {and}
  \bibinfo{person}{Catherine Havasi}.} \bibinfo{year}{2012}\natexlab{}.
\newblock \showarticletitle{Representing General Relational Knowledge in
  ConceptNet 5}. In \bibinfo{booktitle}{{\em Proceedings of the Eighth
  International Conference on Language Resources and Evaluation (LREC-2012)}}.
  \bibinfo{publisher}{European Language Resources Association (ELRA)},
  \bibinfo{address}{Istanbul, Turkey}.
\newblock


\bibitem[\protect\citeauthoryear{Spink, Wolfram, Jansen, and Saracevic}{Spink
  et~al\mbox{.}}{2001}]%
        {spink2001searching}
\bibfield{author}{\bibinfo{person}{Amanda Spink}, \bibinfo{person}{Dietmar
  Wolfram}, \bibinfo{person}{Major B~J Jansen}, {and} \bibinfo{person}{Tefko
  Saracevic}.} \bibinfo{year}{2001}\natexlab{}.
\newblock \showarticletitle{{Searching the web: The public and their queries}}.
\newblock \bibinfo{journal}{{\em Journal of the American Society for
  Information Science and Technology\/}} \bibinfo{volume}{52},
  \bibinfo{number}{3} (\bibinfo{year}{2001}), \bibinfo{pages}{226--234}.
\newblock


\bibitem[\protect\citeauthoryear{Targett}{Targett}{2015}]%
        {Targett2015a}
\bibfield{author}{\bibinfo{person}{Cassandra Targett}.}
  \bibinfo{year}{2015}\natexlab{}.
\newblock \bibinfo{title}{{Apache Solr Reference guide. A quick overview}}.
\newblock   (\bibinfo{year}{2015}).
\newblock
\showURL{%
\url{https://cwiki.apache.org/confluence/display/solr/A+Quick+Overview}}


\bibitem[\protect\citeauthoryear{Xiao, He, Chi, Jeng, and Tomer}{Xiao
  et~al\mbox{.}}{2019}]%
        {Xiao2019}
\bibfield{author}{\bibinfo{person}{Fanghui Xiao}, \bibinfo{person}{Daqing He},
  \bibinfo{person}{Yu Chi}, \bibinfo{person}{Wei Jeng}, {and}
  \bibinfo{person}{Christinger Tomer}.} \bibinfo{year}{2019}\natexlab{}.
\newblock \showarticletitle{{Challenges and supports for accessing open
  government datasets}}. In \bibinfo{booktitle}{{\em Proceedings of the 2019
  Conference on Human Information Interaction and Retrieval - CHIIR '19}},
  \bibfield{editor}{\bibinfo{person}{Leif Azzopardi}, \bibinfo{person}{Martin
  Halvey}, \bibinfo{person}{Ian Ruthven}, \bibinfo{person}{Hideo Joho},
  \bibinfo{person}{Vanessa Murdock}, {and} \bibinfo{person}{Pernilla
  Qvarfordt}} (Eds.). \bibinfo{publisher}{ACM Press},
  \bibinfo{address}{Glasgow, Scotland, United Kingdom},
  \bibinfo{pages}{313--317}.
\newblock
\showISBNx{9781450360258}
\showDOI{%
\url{https://doi.org/10.1145/3295750.3298958}}


\bibitem[\protect\citeauthoryear{Zuiderwijk, Janssen, and Susha}{Zuiderwijk
  et~al\mbox{.}}{2016}]%
        {Zuiderwijk2016a}
\bibfield{author}{\bibinfo{person}{Anneke Zuiderwijk}, \bibinfo{person}{Marijn
  Janssen}, {and} \bibinfo{person}{Iryna Susha}.}
  \bibinfo{year}{2016}\natexlab{}.
\newblock \showarticletitle{{Improving the speed and ease of open data use
  through metadata, interaction mechanisms, and quality indicators}}.
\newblock \bibinfo{journal}{{\em Journal of Organizational Computing and
  Electronic Commerce\/}} \bibinfo{volume}{26}, \bibinfo{number}{1-2}
  (\bibinfo{date}{apr} \bibinfo{year}{2016}), \bibinfo{pages}{116--146}.
\newblock
\showISSN{1091-9392}
\showDOI{%
\url{https://doi.org/10.1080/10919392.2015.1125180}}


\end{thebibliography}

\end{document}